\documentclass[aps,prx,twocolumn,superscriptaddress]{revtex4-2}
\usepackage{amsmath}
\usepackage{amsfonts}
\usepackage{cancel}
\usepackage{amssymb}
\usepackage{graphicx}
\usepackage[colorlinks=True,citecolor=myblue,linkcolor=myorange,urlcolor=mygreen]{hyperref}
\usepackage{hypcap}
\usepackage{braket}
\usepackage{bm}
\usepackage{bbm}
\usepackage{tikz}
\usepackage{xcolor}
\usepackage[export]{adjustbox}
\usepackage{xcolor}

\definecolor{myblue}{RGB}{0,0,130}
\definecolor{myorange}{RGB}{130,50,0}
\definecolor{mygreen}{RGB}{0,130,0}

\usepackage{tikz}
\usetikzlibrary{arrows,decorations.pathmorphing}

\begin{document}

\title{Measurements conspire nonlocally to restructure critical quantum states}
\author{Samuel J. Garratt}
\affiliation{Department of Physics, University of California, Berkeley, California 94720, USA}
\author{Zack Weinstein}
\affiliation{Department of Physics, University of California, Berkeley, California 94720, USA}
\author{Ehud Altman}
\affiliation{Department of Physics, University of California, Berkeley, California 94720, USA}
\affiliation{Materials Science Division, Lawrence Berkeley National Laboratory, Berkeley, California 94720, USA}
\date{\today}
             
\begin{abstract}
We study theoretically how local measurements performed on critical quantum ground states affect long-distance correlations. These states are highly entangled and feature algebraic correlations between local observables. As a consequence, local measurements can have highly nonlocal effects. Our focus is on Tomonaga-Luttinger liquid (TLL) ground states, a continuous family of critical states in one dimension whose structure is parameterized by a Luttinger parameter $K$. We show that arbitrarily weak local measurements, performed over extended regions of space, can conspire to drive transitions in long-distance correlations. Conditioning first on a particular measurement outcome we show that there is a transition in the character of the post-measurement quantum state for $K<1$, and highlight a formal analogy with the effect of a static defect on transport through a TLL. To investigate the full ensemble of measurement outcomes we consider averages of physical quantities which are necessarily nonlinear in the system density matrix. We show how their behavior can be understood within a replica field theory, and for the measurements that we consider we find that the symmetry of the theory under exchange of replicas is broken for $K<1/2$. A well-known barrier to experimentally observing the collective effects of multiple measurements has been the need to postselect on random outcomes. Here we resolve this problem by introducing cross-correlations between experimental measurement results and classical simulations, which act as resource-efficient probes of the transition. The phenomena we discuss are, moreover, robust to local decoherence.
\end{abstract}

\maketitle

\section{Introduction}\label{sec:intro}

Measurements can have nontrivial effects on many-body quantum states. Although collapse is often associated with the loss of quantum correlations, rich new structures can also arise. Indeed, a curious feature of quantum mechanics is the nonlocality of the measurement process, which has striking manifestations in the violation of Bell inequalities \cite{bell1964einstein,aspect1982experimental}, and in the teleportation of quantum information \cite{bennett1993teleporting,bouwmeester1997experimental}. In many-body systems, measurements can furthermore be exploited to perform quantum computation \cite{persistent2001briegel,briegel2009measurement}, highlighting the complexity of the states that one can generate. The loss and generation of quantum correlations through measurement is particularly interesting when the quantum state is, in the first instance, highly entangled.

At low energies, long-range entanglement can arise naturally in the presence of strong quantum fluctuations. Key examples are at quantum phase transitions \cite{sachdev2011quantum} and in one-dimensional quantum liquids \cite{giamarchi2003quantum}, where ground states are critical. In this setting there are algebraic correlations between local observables, and as a consequence a measurement of one of them can modify the expectation values of many others. This behavior should be contrasted with that in thermal states \cite{deutsch2018eigenstate}, which resemble random vectors. Although these states feature extensive entanglement entropies \cite{page1993average}, measuring a single local observable reveals almost no information about any of the others. The information is instead encoded in nonlocal correlations between observables, and so is inaccessible to a conventional observer. 

The nonlocal effects of a single local measurement raise questions over the effects of many. In this work we ask whether measurements performed in different locations in space can conspire with one another to qualitatively alter physical correlations in a quantum state. Focusing on a family of critical ground states, we show how these effects can be described using standard tools from quantum statistical mechanics. Our central result is to show that, for local measurements performed with a finite density in space, there are transitions between phases in which the effects of the measurements are in the one case negligible, and in the other dramatic. Note that these phenomena require that the observer keeps track of the measurement outcomes, since otherwise there can be no teleportation of information.

The critical states we study are described by the theory of Tomonaga-Luttinger liquids (TLLs) \cite{tomonaga1950remarks,luttinger1963exactly,haldane1981effective,haldane1981luttinger,giamarchi2003quantum}. This theory captures the long-wavelength behavior of one-dimensional quantum liquids, both fermionic and bosonic, in terms of density and phase fluctuations. The algebraic correlations in TLLs are highly universal, and for particles without spin they are characterized by a single Luttinger parameter $K$. Smaller values of $K$ correspond to a slower decay of density correlations, and a faster decay of phase correlations. For example, $K=1$ for free fermions, while $K<1$ and $K>1$ describe fermions with repulsive and attractive interactions, respectively. Behavior characteristic of TLLs has been studied experimentally in a wide variety of systems \cite{schwartz1998onchain,yao1999carbon,ruegg2008thermodynamics,lake2005quantum,kono2015field} including ultracold quantum gases \cite{paredes2004tonks,kinoshita2004observation,hofferberth2008probing,yang2017quantum}, where it is possible to probe physical correlations post-measurement~\cite{patil2014nondestructive,patil2015measurement}. 

First, we study the structure of the quantum state prepared by a particular set of weak measurement outcomes. Allowing an ancillary qubit to weakly interact with the local particle density, and subsequently measuring the qubit, there are two possible results: a `click' corresponds to a projective measurement in which we observe a particle, while `no click' only suppresses the amplitude for there to be a particle at the location of the measurement. If there is no click, the particle density remains uncertain. Using this detection scheme at spatial locations commensurate with the mean interparticle spacing, and postselecting for the outcome where there are no clicks, we weakly imprint a charge density wave (CDW) on the quantum state. A perturbative RG analysis reveals that for $K<1$ and for arbitrarily weak measurements there is a transition in the asymptotic form of algebraic correlations in our postselected state. For $K>1$ and for anything short of projective measurement, algebraic correlations characteristic of the unmeasured state persist at long wavelengths. Interestingly, aspects of this problem map onto the study by Kane and Fisher (KF) \cite{kane1992transport,kane1992transmission} of an isolated defect in a TLL. In that problem one finds that at low frequencies the defect causes the system to become insulating for $K<1$, whereas it has a negligible effect for $K>1$.

\begin{figure}
\includegraphics[width=0.47\textwidth]{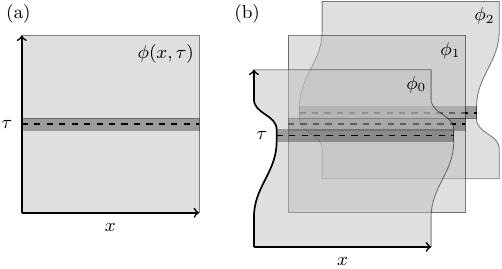}        
\caption{The two classes of problem that we consider in this work. Light grey regions represent the partition function $\text{Tr}e^{-\beta \hat H}$ for Hamiltonian $\hat H$ as an imaginary-time path integral over configurations of a scalar field $\phi(x,\tau)$. Here $x$ denotes the spatial coordinates and $\tau$ the imaginary time. Our focus is on the ground state, so we send $\beta \to \infty$ and restrict our attention to correlations in the surface $\tau=0$. Measurements of observables in the ground state can be represented by operators $\hat M_m$, and these correspond to perturbations at $\tau=0$. (a) For any individual set of measurement outcomes $m$, this leads us to a theory describing a scalar field $\phi(x,\tau)$ in the presence of the perturbation originating from $\hat M_m$ at $\tau=0$. (b) To average over the ensemble of measurement outcomes we use a replica trick. This leads us to a theory describing the fluctuations of a set of $N$ replica fields $\phi_{\alpha}(x,\tau)$ with $\alpha=0, \ldots, (N-1)$, and the average over measurements couples fluctuations $\phi_{\alpha}(x,0)$ for different replica index $\alpha$.}
\label{fig:intro}
\end{figure}

To characterize fully the effects of density measurements on a quantum state, we then consider an entire ensemble of measurement outcomes. For analytic simplicity, here we use as our ancillary degree of freedom an oscillator rather than a qubit. If we average physical quantities over the ensemble of outcomes, and weight the various contributions by Born probabilities, then to characterize the influence of our measurements we must work with quantities that are nonlinear in the state. This is necessary because averaging physical quantities linear in the state and weighting the results by Born probabilities is equivalent to averaging the state itself, which corresponds to ignorance of the measurement outcomes. One can consider, for example, squared density correlations averaged over measurement outcomes. To calculate averages of nonlinear correlation functions such as these we formulate a replica field theory. At long wavelengths our theory is closely related to (but is distinct from) that in the KF problem, and here also we find a transition in the behavior of correlation functions. This transition corresponds to a spontaneous breaking of the symmetry associated with the exchange of replicas, and occurs for $K<1/2$. 

Finally, we discuss how phenomena occurring in the ensemble of post-measurement states can be observed in experiment without the need to postselect on random measurement outcomes. Our approach requires that post-measurement expectation values of local observables can also be estimated on a classical computer; for the critical states of interest here, the computational resource requirements for such a calculation are only polynomial in system size \cite{verstraete2006matrix}. The basic idea is to cross-correlate results of classical simulations with data from experiment, and to average these cross-correlations over experimental runs. This average washes out quantum fluctuations, and allows one to estimate measurement-averaged quantities that are nonlinear in the quantum state. We note that the idea of using classical computers to avoid a postselection overhead was previously discussed in Ref.~\cite{gullans2020scalable,dehghani2022neural,li2021robust}. A separate concern in experiment is environmental decoherence. However, in the absence of any subsequent dynamics, local quantum channels can only alter the expectation values of observables with which they share support. Few-body correlation functions are therefore a robust probe of the effects of measurement.

A related class of problems concern the effects of measurements on many-body systems that are additionally evolving under unitary dynamics \cite{li2018quantum,skinner2019measurement,chan2019unitary,cao2019entanglement}, and we note that the consequences of nonlocality have there been investigated in Refs.~\cite{li2021conformal,friedman2022locality}. Critical quantum states are moreover abundant in that setting, and this fact provides additional motivation for understanding how they respond to measurement. For example, when the unitary dynamics is chaotic, there is a threshold measurement rate at which the system undergoes a dynamical measurement-induced phase transition (MIPT) \cite{li2018quantum,skinner2019measurement} separating volume-law and area-law entangled states. At the transition itself, the system evolves within an ensemble of critical states \cite{li2018quantum,skinner2019measurement,li2019measurement,bao2020theory,jian2020measurement,turkeshi2020measurement}. Meanwhile, in systems of noninteracting fermions, there appears to be critical behavior at low measurement rates \cite{alberton2021entanglement,turkeshi2021measurement,bao2021symmetry,buchhold2021effective,muller2022measurement,minoguchi2022continuous}. Strikingly, ensembles of critical states arise even under dynamics that consists solely of measurements that do not commute \cite{lavasani2021measurement,ippoliti2021entanglement,regemortel2021entanglement}. 

In the static setting of interest here, related works have considered the characterization \cite{benzion2020disentangling} and creation \cite{piroli2021quantum,verresen2021efficiently,tantivasadakarn2021longrange,lin2022probing,lu2022measurement} of entangled states using measurement. The measurement-induced teleportation of information in many-body states has meanwhile been investigated in Refs.~\cite{bao2021finite,lin2022probing}. In particular, the results of Ref.~\cite{bao2021finite} have revealed that in two (or more) spatial dimensions the quantum states prepared by local unitary dynamics undergo a transition in their response to measurement at a finite time. Beyond this time, if the observer performs projective measurements of all but two arbitrarily well-separated degrees of freedom, their resulting quantum state can remain entangled. Focusing on this measurement scheme but instead considering critical states, Ref.~\cite{lin2022probing} has recently shown that the entanglement between the unmeasured degrees of freedom is sensitive to the sign structure of the state. Notably, Refs.~\cite{rajabpour2015post,rajabpour2016entanglement} also studied the effects of measuring a finite region of space on entanglement in a critical state, restricting to the case of a nonrandom set of outcomes. In this work we are instead concerned with weak local measurements of essentially all degrees of freedom. Such measurements extract only partial information on local observables, and all constituents of the system typically remain entangled with one another. Moreover, our focus is primarily on the full ensemble of measurement outcomes.

This paper is organized as follows. First, in Sec.~\ref{sec:overview}, we provide overviews of the problems we consider and of our results. In Sec.~\ref{sec:qnd} we then discuss the state resulting from a particular set of weak measurement outcomes. Following this, in Sec.~\ref{sec:main} we consider averages over an ensemble of outcomes. In Sec.~\ref{sec:postselection} we discuss the postselection problem, and how it can be avoided. We provide a summary, and indicate outstanding questions, in Sec.~\ref{sec:discussion}.

\section{Overview}\label{sec:overview}

The basic structure of the problem is as follows. Starting from a ground state $\ket{\psi_{\text{g.s.}}}$ of a Hamiltonian $\hat H$, we consider performing an extensive number of weak local measurements. Physically, we imagine introducing ancillary degrees of freedom, and allowing them to briefly interact with the system. Subsequent projective measurements of the ancillae give rise to a nonunitary update of the state of the system. These weak measurements alter the amplitudes of the various contributions to the many-body state, but do not fully disentangle the system degrees of freedom from one another. Consequently, the measured state is still highly nontrivial. We are interested in whether the asymptotic properties of correlation functions are modified relative to the ground state. 

In Sec.~\ref{sec:overviewfield} we outline how this situation can be described within a Euclidean field theory. The measurements appear as a kind of randomness in this theory, and in Sec.~\ref{sec:overviewaverage} we discuss how to treat this feature of the problem. The specific systems that we focus on in this work are described in Sec.~\ref{sec:overviewTLL}, and our theoretical results are summarised in Sec.~\ref{sec:overviewresults}.

\subsection{Field theory}\label{sec:overviewfield}
It is convenient to express the projector onto the ground state $\ket{\psi_{\text{g.s.}}}\bra{\psi_{\text{g.s.}}}$ as imaginary-time evolution $e^{-\beta \hat H}$ with $\beta \to \infty$. Let us write this imaginary-time evolution as a path integral in the basis of eigenstates of a Hermitian quantum field $\hat \phi(x)$. For example, the partition function $\text{Tr} e^{-\beta \hat H} = \int D \phi \, e^{-S[\phi]}$. Here $\phi=\phi(x,\tau)$ is a scalar field of eigenvalues of $\hat \phi(x)$, the action $S[\phi]$ is an integral over spatial coordinates $x$ and the imaginary time $\tau$, and in the partition function the boundary conditions are $\phi(x,0)=\phi(x,\beta)$. For a $d$-dimensional quantum system, the structure of the ground state is encoded in equal-$\tau$ correlation functions in this $(d+1)$-dimensional field theory.

For weak measurement outcomes that we denote by $m$ the state after measurement is ${\ket{\psi_m} = \hat M_m \ket{\psi_{\text{g.s.}}} / p^{1/2}_m}$, where $\hat M_m$ is a nonunitary Hermitian operator. The normalization is set by the Born probability $p_m = \braket{\hat M^2_m}_{\text{g.s.}}$, and $\braket{\ldots}_{\text{g.s.}}$ denotes an expectation value in state $\ket{\psi_{\text{g.s.}}}$. We only require that the set of $\hat M_m$ corresponding to the different outcomes $m$ satisfies the probability-conserving condition $\sum_m \hat M^2_m=1$ (i.e. that together they form a quantum channel). Correlation functions in the measured state $\ket{\psi_m}$ are computed from its density matrix
\begin{align}
    \ket{\psi_m}\bra{\psi_m} = \lim_{\beta \to \infty} \frac{\hat M_m e^{-\beta \hat{H}} \hat M_m}{\text{Tr}[ \hat M^2_m e^{-\beta \hat H}]}.
\end{align}
Expectation values $\braket{\ldots}_m$ in the state $\ket{\psi_m}$ are then given by
\begin{align}
    \braket{\ldots}_m = \frac{\int D\phi \, \braket{\varphi'| \hat M_m (\ldots) \hat M_m|\varphi} e^{-S[\phi]} } {\int D\phi\braket{\varphi'|\hat M^2_m|\varphi} e^{-S[\phi]}},
\end{align}
where for brevity we have defined the fields $\varphi(x) \equiv \phi(x,0)$ and $\varphi'(x) \equiv \phi(x,\beta)$. In the case where $\hat M_m$ and the observable of interest (here represented by the ellipsis) commute with $\hat \phi(x)$ we have $\varphi = \varphi'$, so the measurement $\hat M_m$ can be viewed as acting at a fixed imaginary time which we have chosen to be $\tau=0$. For $\hat M_m$ that acts throughout space, the measurements then appear as perturbations on the $d$-dimensional $\tau=0$ surface in the $(d+1)$-dimensional field theory, and this construction is illustrated in Fig.~\ref{fig:intro}(a). Questions about the asymptotic properties of correlation functions in the state immediately following measurement are then questions about whether this perturbation alters correlations within the $\tau=0$ surface. For critical quantum ground states that correspond to RG fixed points, we must ask whether the perturbation representing the measurement is relevant in the appropriate fixed-point theory.

\subsection{Averaging}\label{sec:overviewaverage}

In Secs.~\ref{sec:qnd} and \ref{sec:main} we approach this problem in two different ways. In Sec.~\ref{sec:qnd} we consider the quantum state arising from one set of measurement outcomes. The outcomes that we choose correspond to a perturbation on the $\tau=0$ surface that is invariant under spatial translations. In this case there is analytic simplicity, as well as an interesting connection to equilibrium behavior in the presence of static defects. More generally, however, we must consider physical quantities averaged over the ensemble of measurement outcomes, and this is the focus of Sec.~\ref{sec:main}.

It is essential that the quantities we average are nonlinear in the density matrices $\ket{\psi_m}\bra{\psi_m}$. This is because averaging $\ket{\psi_m}\bra{\psi_m}$ with weights given by the Born probabilities $p_m$ is equivalent to dephasing in the basis of eigenstates of the measured operators, and dephasing events do not have nonlocal effects on the expectation values of observables. In order to calculate averages of nonlinear quantities, such as squared correlation functions $\braket{\ldots}_m^2$, we develop a replica field theory. This comes from first writing e.g.
\begin{align}
	\sum_m \, p_m \braket{\ldots}_m^2 = \lim_{N \to 1} \frac{\sum_m \, p^N_m \braket{\ldots}_m^2}{\sum_m \, p^N_m}. \label{eq:overviewreplica}
\end{align} 
Here the different possible measurement outcomes will correspond to different configurations of a scalar field $m(x)$, so the sum $\sum_m$ should be interpreted as an integral. The above trick allows us to make analytic progress for integer $N \geq 2$. Physically, performing calculations for $N > 1$ corresponds to overemphasizing contributions from the most likely measurement outcomes. 

To arrive at the replica field theory we write each of $p_m$ and $\braket{\ldots}_m$ in terms of the path-integral representation of $e^{-\beta \hat H}$. Averages of nonlinear correlation functions become $\tau=0$ correlations in a theory of $N$ replica fields $\phi_{\alpha}(x,\tau)$, where the replica index $\alpha = 0 \ldots (N-1)$.  In this theory, each field $\phi_{\alpha}$ interacts with the same $\tau=0$ perturbation corresponding to $\hat M_m$. Averaging over measurement outcomes has the effect of weakly `locking' the replicas together at $\tau=0$, as shown in Fig.~\ref{fig:intro}(b). This locking has a physical interpretation as the suppression of quantum fluctuations of measured observables. 

\subsection{Tomonaga-Luttinger liquids}\label{sec:overviewTLL}

While the framework described above is much more general, in this paper we focus on critical states in $d=1$ described by the theory of TLLs \cite{giamarchi2003quantum,haldane1981luttinger}, and for simplicity we consider spinless fermions. We measure the particle density $\hat n(x)$, which can be expressed in terms of a counting field $\hat \phi(x)$ that describes the displacement of particles from a putative ordered lattice arrangement. Explicitly, the normal-ordered density operator is
\begin{align}
 	\hat n(x) = -\pi^{-1} \nabla \hat \phi(x) + \pi^{-1} \cos [2 (k_F x - \hat \phi(x))],
\label{eq:density}
\end{align}
where we have fixed the microscopic length scale in the problem to unity. The wavenumber $k_F$ sets the mean interparticle separation $\pi/k_F$, and we have neglected contributions to $\hat n(x)$ oscillating with wavenumber $4k_F, 6k_F, \ldots$ since these do not affect our results. In this setting the counting field will play the role of the general quantum field $\hat \phi(x)$ discussed earlier in this section. Note also that in an infinite system the theory is symmetric under shifts of the counting field by $\pi$. 

The long-wavelength form of the action for a TLL, appearing in the path-integral representation of $\text{Tr}e^{-\beta \hat H}$, is in the density representation given by
\begin{align}
	S[\phi] =  \frac{1}{2\pi K} \int dx \int_0^{\beta} d\tau [ \dot \phi^2 + (\nabla \phi)^2 ],
	\label{eq:SLL}
\end{align}
where $\dot \phi$ and $\nabla \phi$ are derivatives of the real scalar field $\phi(x,\tau)$ with respect to $\tau$ and $x$, respectively. Correlations of the phase $\hat \theta(x)$ follow from rewriting this action using the canonical commutation relation $[\hat \phi(x),\nabla \hat \theta(x')] = i\pi \delta(x'-x)$. The action describing phase fluctuations has the same form as $S[\phi]$, but with the role of $K$ replaced by $K^{-1}$. Smaller values of $K$ correspond to stronger density correlations and weaker phase correlations.

\subsection{Results}\label{sec:overviewresults}
In this work we show that there are transitions, occurring as a function of $K$, in the effects that measurements have on the ground states of TLLs. To illustrate the idea in Sec.~\ref{sec:qnd} we consider weak measurements of the local density, using ancillary qubits, at locations commensurate with the mean interparticle separation. For a set of measurement outcomes where no particles are detected with certainty, which we refer to as `no clicks', we weakly imprint a CDW on the many-body state. At the level of the field theory, this corresponds to a perturbation of the form $\delta S \propto \int dx \cos [2 \phi(x,0)]$ added to the action Eq.~\eqref{eq:SLL}. This action is equivalent, following an exchange of time and space coordinates, to the one used to describe a local defect in a TLL in the KF problem \cite{kane1992transport} [see Fig.~\ref{fig:KF}].

Just as in the static defect problem, we can examine the scaling of the no-click perturbation under RG with the same result. The perturbation is relevant for $K<1$ and irrelevant for $K>1$, which implies a transition in the structure of the measured state $\ket{\psi_{\text{n.c.}}}$ at the critical value of the Luttinger parameter $K=1$. The transition involves a change in the exponents governing power-law decays of correlation functions. For $K>1$,  the asymptotic decay of phase correlations conditioned on observing no clicks is unchanged relative to the ground state, i.e. $\braket{e^{i[\hat \theta(x) - \hat \theta(0)]}}_{\text{n.c.}} \sim x^{-1/(2K)}$, where $\braket{\ldots}_{\text{n.c.}}$ denotes an expectation value with respect to $\ket{\psi_{\text{n.c.}}}$. On the other hand, for $K<1$ the asymptotic phase correlations change to $\braket{e^{i[\hat \theta(x) - \hat \theta(0)]}}_{\text{n.c.}} \sim x^{-1/K}$. Thus, we find that in the first case ($K>1$) anything short of a projective measurement fails to alter the asymptotic behavior of correlation functions, in the second ($K<1$) an arbitrarily weak measurement causes a strong suppression of phase correlations at long distances.

In Sec.~\ref{sec:main} we consider physical quantities averaged over the ensemble of all measurement outcomes. For this purpose it is useful to consider ancillary oscillators instead of the qubits used in Sec. \ref{sec:qnd}. This choice of measurement scheme allows us to perform the average over outcomes analytically using a replica trick, which introduces a set of $N$ replica fields $\phi_{\alpha}$. In the limit of vanishing coupling $\mu$ between system and ancillae, each of the fields $\phi_{\alpha}$ is independently described by the action $S[\phi_{\alpha}]$. With nonvanishing coupling, the average over measurements generates a perturbation in the replica field theory, which couples the replica fields at $\tau=0$: 
\begin{equation}
\delta S \propto -\mu\sum_{\alpha < \beta}\int d x \cos[2(\phi_{\alpha}(x,0)-\phi_{\beta}(x,0))].  
\end{equation}
This perturbation favours field configurations in which the replicas are locked together. 

We show that, for weak coupling (small $\mu$) between system and ancillae, the measurement-induced locking of replicas is a relevant perturbation for Luttinger parameter $K<1/2$, and that it is irrelevant for $K>1/2$. This criterion is independent of $N$, suggesting that there is a transition in the behavior of averaged nonlinear correlation functions even for $N \to 1$, i.e. when the contributions from different measurement outcomes are weighted by the Born probabilities $p_m$. For strong coupling between system and ancillae, we are able to show that for $N=2$ the critical Luttinger parameter remains $K=1/2$. The transition at $K=1/2$ has signatures in the power-law decays of averaged nonlinear correlation functions; for $K<1/2$ the density measurements conspire to suppress quantum fluctuations of the density, and correlations of the phase. 

In Sec.~\ref{sec:postselection} we show how this transition can be observed without the need to postselect on random measurement outcomes, and so with modest experimental resources. To do this we introduce as probes of the transition cross-correlations between measurement results and classical simulations. Provided it is possible to calculate the conditional expectation values $\braket{\ldots}_m$ of interest, these probes allow one to estimate physical quantities having, for example, the structure of the left-hand side of Eq.~\eqref{eq:overviewreplica}.

\begin{figure}
\includegraphics[width=0.47\textwidth]{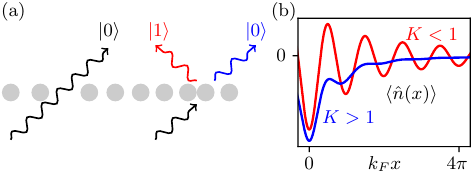}   
\caption{(a) Ancilla qubits initially in state $\ket{0}$ interact weakly with the local particle density, and are then measured in the computational basis. If there is no particle, the result of the measurement is $\ket{0}$. If there is a particle, the result is either $\ket{0}$ or $\ket{1}$. Therefore, if we record outcome $\ket{0}$, i.e. `no click', the particle density remains uncertain. Here we represent the states $\ket{0}$ and $\ket{1}$ as a pair of photon modes. (b) Illustration of the density profile $\braket{\hat{n}(x)}$ following no click at $x=0$ for (red) $K<1$ and (blue) $K>1$.}
\label{fig:QND}
\end{figure}

\section{No clicks}\label{sec:qnd}

In this section we consider weak measurements of local densities using ancillary qubits and postselect for a particular set of outcomes. We are therefore concerned with correlations in a single quantum state. It is simplest to first discuss the protocol at a single site. We initialize a qubit (representing our measurement apparatus) in eigenstate $\ket{0}$ of a Pauli $\hat \sigma^z$ operator, and then couple it to the density $\hat n_j$ at a site $j$. Following this, we projectively measure $\hat \sigma^z$. Note that the ancilla need not correspond to a physical qubit, and in practice there are a variety of physical implementations of this protocol. For example, we could also imagine scattering photons off of the system, and detecting whether they are scattered from an initial mode $\ket{0}$ to another, $\ket{1}$, as illustrated in Fig.~\ref{fig:QND}(a). 

The coupling between $\hat n_j$ and the ancilla is as follows. First note that here the lattice operator $\hat n_j$ is not normal ordered, while the continuum operator $\hat n(x)$ is. In other words, $\hat n(x)$ is the continuum analogue of $\hat n_j - \braket{\hat n_j}_{\text{g.s.}}$. The system and ancilla first evolve under the unitary operator
\begin{align}
	\hat U_j &= \exp[ i u_j \hat n_j \otimes  \hat \sigma^x ], \label{eq:QNDU}\\
	&= 1 + [\cos u_j-1]\hat n_j + i[\sin u_j] \hat n_j \otimes \hat\sigma^x \notag
\end{align}
where $u_j$ sets the strength of the coupling. We have simplified the expression by using $\hat n_j^2 = \hat n_j$, which applies for fermions. After acting with $\hat U_j$ we perform a projective measurement of $\hat \sigma^z$, with outcome $\ket{0}$ or $\ket{1}$. The states of the system that result from a single measurement are, respectively,
\begin{align}
	\ket{\psi_0} &= \big( 1 + [\cos u_j-1]\hat n_j \big)\ket{\psi_{\text{g.s.}}}, \label{eq:psi01}\\
	\ket{\psi_1} &= i[\sin u_j]\hat n_j \ket{\psi_{\text{g.s.}}}, \notag
\end{align}
These states are not normalized, and for example the probability for result $\ket{0}$ is $\braket{\psi_0|\psi_0}$. Note that for outcome $\ket{1}$, which we refer to as a `click', there is a fermion at site $j$ \textit{with certainty}. This is due to the appearance of the projection operator $\hat n_j$ in the expression for $\ket{\psi_1}$. For outcome $\ket{0}$, or `no click', we do not know whether there is a fermion at $j$ (unless $u_j=\pi/2$, since this corresponds to a standard projective measurement of $\hat n_j$). However, the expectation value of $\hat n_j$ is in general suppressed relative to $\braket{\hat n_j}_{\text{g.s.}}$. 

Before proceeding, it is helpful to develop some intuition for the effects of these weak measurements, and for the role of $K$. For smaller $K$ there are stronger density correlations in the ground state, with the oscillatory contribution to the density in Eq.~\eqref{eq:density} decaying as $x^{-2K}$. The effects of individual measurements of the density are therefore felt out to greater distances. This is illustrated in Fig.~\ref{fig:QND}(b), where we show the nonlocal effects of a single no-click outcome on the density profile for two different values of $K$. 

This fact provides a hint as to the behavior we can expect when many measurements are performed. As a first demonstration, in this section we apply the above measurement protocol to an extended region of space and postselect for the case where no clicks are observed. The resulting state is
\begin{align}
    	\ket{\psi_{\text{n.c.}}} &= \braket{\hat M^2_{\text{n.c.}}}^{-1/2}_{\text{g.s.}} \hat M_{\text{n.c.}} \ket{\psi_{\text{g.s.}}}. \notag\\
	\hat M_{\text{n.c.}} &\equiv \prod_{j} \big( 1 + [\cos u_j-1]\hat n_j \big),
\label{eq:psinc}
\end{align}
and throughout this section we will consider the structure of correlation functions in $\ket{\psi_{\text{n.c.}}}$. This problem is simplified considerably for weak measurements (as opposed to projective ones) since the classical information extracted decreases continuously with $u_j$, and for small $u_j$ we can consider the effect of extracting this information in perturbation theory.

\subsection{Field theory}

Here we formulate the problem of evaluating density correlations in $\ket{\psi_{\text{n.c.}}}$ in terms of the field theory outlined in Sec.~\ref{sec:overview}. It is convenient to write ${e^{-(v_j/2) \hat n_j} = 1 + [\cos u_j-1]\hat n_j}$, i.e. $v_j = -2\ln |\cos u_j|$. Note that for $u_j=\pi/2$, corresponding to a projective measurement, we have infinite $v_j$. For weak measurements we instead have ${v_j = u^2_j + O(u_j^4)}$. The effects of our measurements on the state are described by
\begin{align}
	\hat M_{\text{n.c.}} = e^{- \frac{1}{2}\sum_j v_j  \hat n_j} \propto e^{- \frac{1}{2}\int dx v(x) \hat n(x)},
\end{align}
where we have switched to continuum notation, and have omitted a constant prefactor arising from the fact that $\hat n(x)$ is normal ordered whereas $\hat n_j$ is not.  

The density correlations in $\ket{\psi_{\text{n.c.}}}$ are given by
\begin{align}
	\braket{\hat n(0) \hat n(x)}_{\text{n.c.}} &= \frac{\text{Tr} \big[ e^{-\beta \hat H} \hat M^2_{\text{n.c.}} \hat n(0) \hat n(x)\big]}{\text{Tr} \big[ e^{-\beta \hat H} \hat M_{\text{n.c.}}^2 \big]}.
\label{eq:densitydensity}
\end{align} 
where we have used $[\hat n(x),\hat M_{\text{n.c.}}]=0$. We can write this correlator in terms of path integrals over $\phi$. Furthermore, since we are interested in correlations at $\tau=0$, we integrate out fluctuations of the field $\phi(x,\tau)$ at $\tau \neq 0$. Writing $\varphi(x)=\phi(x,\tau=0)$, and taking the Fourier transform $\tilde \varphi(q) = \int dx \, e^{-iqx} \varphi(x)$, this integration gives the nonlocal action
\begin{align}
	s[\varphi] = \frac{1}{\pi K} \int \frac{dq}{2\pi} |q| |\tilde \varphi(q)|^2.
\label{eq:smalls}
\end{align}
It can be verified that Eq.~\eqref{eq:smalls} generates the same $\tau=0$ correlations as $S[\phi]$. Note that the inverse Green's function $\sim |q|$ corresponds to interactions decaying as $\sim x^{-2}$ in real space.

Within this formulation we can write the numerator in Eq.~\eqref{eq:densitydensity} as
\begin{align}
	\int D\varphi \, e^{-s[\varphi] - \int dx' v(x') n(x')} n(0) n(x),
\end{align}
with $s[\varphi]$ given in Eq.~\eqref{eq:smalls}. The scalar field $n(x)$ is a function of $\varphi(x)$ given by replacing $\hat \phi(x)$ with $\varphi(x)$ on the right-hand side of Eq.~\eqref{eq:density}. The perturbation $\int dx \, v(x) n(x)$ due to our measurements depends sensitively on the form of $v(x)$. As we will see, interesting effects arise from the component oscillating with wavenumber $2k_F$. To highlight these effects we can imagine performing weak measurements at locations commensurate with the mean interparticle spacing. Density correlations in $\ket{\psi_{\text{n.c.}}}$ are then evaluated as averages with respect to the action
\begin{align}
	s_{\text{n.c.}}[\varphi] \equiv s[\varphi] - v \int dx \cos[2\varphi],
\label{eq:snc}
\end{align}
where $v$ is proportional to the $2k_F$ Fourier component of $v(x)$.  Observing no clicks then has the effect of weakly pinning the field $\varphi$ to an integer multiple of $\pi$. In the next section we will determine when this effect alters the long-wavelength structure of correlation functions.

\begin{figure}
\includegraphics[width=0.47\textwidth]{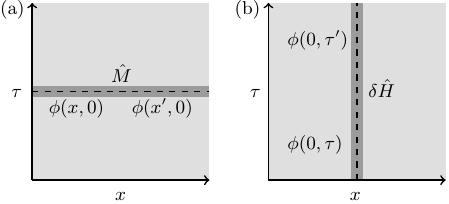}    
\caption{Relation to transport through a defect in a TLL. (a) In this work we are interested in quantum states perturbed by measurements $\hat M$ at all locations in space $x$, and their effect on correlations in $x$ at fixed $\tau$. (b) In the defect problem one is interested in a spatially local and time-independent perturbation $\delta \hat H$ to the Hamiltonian, and its effect on correlations in $\tau$ at fixed $x$.}
\label{fig:KF}
\end{figure}

Equation~\eqref{eq:smalls} is equivalent to the action for a local degree of freedom coupled to a zero-temperature Ohmic bath \cite{caldeira1981influence}, although with the roles of $x$ and $\tau$ interchanged. Integrating out a physical bath, corresponding for example to linearly-dispersing bosons at $x \neq 0$, generates an action for a local degree of freedom at $x=0$ that is nonlocal in (imaginary) time. This temporal nonlocality encodes the memory of the bath. In our case the `bath' is replaced by fluctuations of the field $\phi(x,\tau)$ at $\tau \neq 0$. The spatial nonlocality of the action Eq.~\eqref{eq:smalls} encodes the entanglement in the ground state.

Moreover, exchanging $x$ and $\tau$ in Eq.~\eqref{eq:snc}, the action is identical to the one generating temporal correlations at a local defect potential in the KF problem~\cite{kane1992transport,kane1992transmission}. This correspondence is illustrated in Fig.~\ref{fig:KF}. In that setting it was shown that such a defect potential is relevant for $K<1$, leading to insulating behavior, but it does not affect the low-frequency conductance for $K>1$. In direct analogy with those results, here we find that long-wavelength correlations in $\ket{\psi_{\text{n.c.}}}$ show striking departures from those in $\ket{\psi_{\text{g.s.}}}$ for $K<1$, while they are essentially unchanged for $K>1$.

\subsection{Transition}\label{sec:nctransition}
Here we identify the regime in which the asymptotic properties of correlations are affected by the measurements discussed above. First note that since we are postselecting for no clicks at locations commensurate with CDW order, the resulting quantum state certainly has $\braket{\cos 2\hat \phi(x)}_{\text{n.c.}} > 0$; by construction, there is long-range CDW order. However, the measurements can have nontrivial effects on correlation functions of the smooth part of the density $\braket{\nabla \hat \phi(0) \nabla \hat \phi(x)}_{\text{n.c.}}$, and of the phase $\braket{e^{i[\hat \theta(x)-\hat \theta(0)]}}_{\text{n.c.}}$. To determine the effect on the long range correlations we apply a standard RG scheme. If the measurement strength flows to zero under RG transformations, then the asymptotics of $\braket{\nabla \hat \phi(0) \nabla \hat \phi(x)}_{\text{n.c.}}$, and of the phase correlations $\braket{e^{i[\hat \theta(x)-\hat \theta(0)]}}_{\text{n.c.}}$, will be unchanged relative to their behavior in $\ket{\psi_{\text{g.s.}}}$. On the other hand, if measurements are relevant, we will see that the powers governing the algebraic decays of these correlation functions are altered.

We first outline the perturbative RG treatment of $s_{\text{n.c.}}[\varphi]$ in Eq.~\eqref{eq:snc} for weak measurements. In this case we expand $e^{-s_{\text{n.c.}}[\varphi]}$ to first order in $v$. With initial UV cutoff $\Lambda$, we write $\tilde \varphi(q) = \tilde \varphi_{<}(q) + \tilde \varphi_{>}(q)$, where $\tilde \varphi(q)=\tilde \varphi_<(q)$ for $|q| < \Lambda e^{-\ell}$ and $\tilde \varphi(q)=\tilde \varphi_>(q)$ for $|q| > \Lambda e^{-\ell}$, integrate out the fields $\tilde \varphi_>(q)$, and rescale lengths $x'=x e^{-\ell}$. If we do not rescale $\varphi(x)$, the first term in Eq.~\eqref{eq:snc} is invariant under the RG, while the parameter $v$ flows as
\begin{align}
	\frac{dv}{d\ell} = (1-K)v.
\label{eq:flowu}
\end{align}
Physically, fluctuations of $\varphi$ on short wavelengths act to minimize the effect of the CDW pinning $\cos 2\varphi$. For ${K<1}$ these fluctuations are small, which is to be expected for fermions with repulsive interactions. As a consequence, $v$ increases under the RG. For $K>1$ on the other hand, there are relatively large density fluctuations on short wavelengths, and so $v$ decreases.

One can similarly carry out RG transformations that are appropriate for strong measurements \cite{furusaki1993single,giamarchi2003quantum}. It is clear that $\cos 2\varphi$ is maximized for $\varphi = p \pi$ for $p$ integer, and that jumps in $p$ correspond to defects in the CDW order. We refer to these defects as domain walls, although from Eq.~\eqref{eq:density} we see that away from them and for any $p$ the density $\sim \cos[2k_Fx]$.
In the limit of large $v$ every configuration of domain walls corresponds to a different saddle-point of the action, and physical correlations are controlled by configurations in which domain walls are dilute, with typical separations much larger than their width as illustrated in Fig.~\ref{fig:RGnc}. In Appendix~\ref{sec:walls} we discuss this saddle-point approximation in detail, following Ref.~\cite{giamarchi2003quantum}. Since we are interested only in correlations on length scales much larger than the domain-wall width, it suffices to approximate
\begin{align}
	\varphi(x) \simeq \pi \sum_j \epsilon_j \Theta(x-x_j),
\label{eq:sharpwalls}
\end{align}
where $\Theta(x)$ is the step function, $x_j$ are locations of domain walls, and $\epsilon_j = \pm 1$ are their signs. Substituting this into Eq.~\eqref{eq:smalls} one finds a logarithmic interaction between domain walls $2K^{-1} \sum_{j < k} \epsilon_j \epsilon_k \log|x_j-x_k|$ such that domain walls with opposite $\epsilon$ attract, while those with the same $\epsilon$ repel. As we coarse-grain in real space, changing the minimum length scale in the problem by a factor $b=e^{\ell}$, we annihilate oppositely-signed domain walls with separation smaller than $b$. When $K$ is sufficiently small, and so the attraction between such domain walls sufficiently strong, this procedure causes domain walls to become ever more dilute. For large $v$ this leads to the RG flow
\begin{align}
	\frac{dv^{-1}}{d\ell} \propto (1-1/K)v^{-3/2},
\label{eq:flowy}
\end{align}
where we have omitted a constant of order unity. Although Eqs.~\eqref{eq:flowu} and \eqref{eq:flowy} are respectively appropriate only for weak and for strong measurements, if we connect together the RG flows we see that for $K<1$ the long-wavelength behavior is described by dilute domain walls while for $K>1$ it is described by the unmeasured theory in Eq.~\eqref{eq:smalls}; there is therefore a transition in the response of the quantum state to measurement at $K=1$. Figure~\ref{fig:RGnc} shows the flow of $v$ as a function of $K$, as well as the behavior of long-wavelength components of the field $\varphi$.

\begin{figure}
\includegraphics[width=0.47\textwidth]{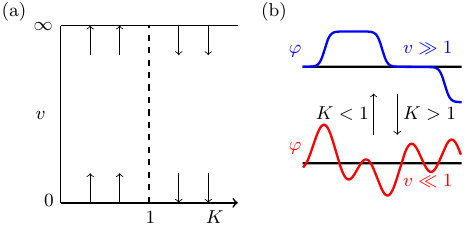}        
\caption{Transition in the structure of the state $\ket{\psi_{\text{n.c.}}}$ as a function of $K$. (a) RG flow of the measurement strength $v$, which changes direction across $K=1$. (b) The character of the field configurations that dominate correlation functions. For $K<1$ long-wavelength components of $\varphi$ are pinned to integer multiples of $\pi$ by $v\cos 2\varphi$, while for $K>1$ they behave as in the unmeasured system. The directions of the arrows on the right indicate the effect of coarse-graining.}
\label{fig:RGnc}
\end{figure}

The transition has dramatic implications for the structure of correlation functions in $\ket{\psi_{\text{n.c.}}}$. For $K>1$, where the measurements are irrelevant, the algebraic decays of correlation functions in $\ket{\psi_{\text{n.c.}}}$ are as in $\ket{\psi_{\text{g.s.}}}$. If our measurement is strong, however, we should only expect to see this behavior on large length scales. For $K<1$ and for an arbitrarily weak measurement, on the largest scales the exponents governing the algebraic decays of correlation functions are modified, as we now discuss.

It is natural to expect that for $K<1$ correlations between density fluctuations $\nabla \hat \phi$ are suppressed relative to $\braket{\nabla \hat \phi(0) \nabla \hat \phi(x)}_{\text{g.s.}} \sim -x^{-2}$ \cite{giamarchi2003quantum}. To see that this is the case we consider the regime of dilute domain walls. There we have $\nabla \varphi(x) \simeq \pi \sum_j \epsilon_j \delta(x-x_j)$ from Eq.~\eqref{eq:sharpwalls}. The long-range interaction between domain walls discussed below Eq.~\eqref{eq:sharpwalls} then leads to
\begin{align}
	\braket{\nabla \hat \phi(0) \nabla \hat \phi(x)}_{\text{n.c.}} \sim -x^{-2/K}, \label{eq:ncgradphi}
\end{align}	
with a prefactor that is set by the square of the rescaled domain wall fugacity. From Eq.~\eqref{eq:ncgradphi} we indeed find that since $K<1$ the $\nabla \hat \phi$ correlation function decays more rapidly than in the unmeasured TLL. Note that if we were to neglect domain walls, and compute $\braket{\nabla \hat \phi(0) \nabla \hat \phi(x)}_{\text{n.c.}}$ by considering quadratic fluctuations around the leading saddle point, i.e. expanding $\cos 2\varphi \simeq 1 - 2\varphi^2$ in Eq.~\eqref{eq:snc}, one would instead find a decay as $x^{-4}$. This is clearly a subleading contribution for $K>1/2$. 

Separately, we can ask about phase correlations. Due to the conjugacy of $\hat \phi$ and $\nabla \hat \theta$ we anticipate that these are weaker than in the unmeasured system, where $\braket{e^{i[\hat \theta(0)-\hat \theta(x)]}}_{\text{g.s.}} \sim x^{-1/(2K)}$. Working with the domain wall picture, in Appendix~\ref{sec:nccorrelations} we show that for $K<1$,
\begin{align}
 	\braket{e^{i[\theta(0)-\theta(x)]}}_{\text{n.c.}} \sim x^{-1/K}. \label{eq:nctheta}
\end{align}
Note that there is an important difference in the calculation of phase correlations relative to density correlations, simply because $\hat \theta(x)$ does not commute with density measurements. The result Eq.~\eqref{eq:nctheta} shows that phase fluctuations are indeed enhanced by our density measurements, and this is reflected in a doubling of the exponent in the correlation function as one decreases $K$ through $K=1$. Again, if we also consider quadratic fluctuations around the leading saddle point, we find only a subleading contribution to Eq.~\eqref{eq:nctheta}; in this case the generation of mass term $\cos 2\varphi \simeq 1 - 2\varphi^2$ causes the subleading contribution to decay exponentially. In Appendix~\ref{sec:numerics} we present numerical calculations of the correlation functions $\braket{\nabla \hat \phi(0) \nabla \hat \phi(x)}_{\text{n.c.}}$ and $\braket{e^{i[\hat \theta(0)-\hat \theta(x)]}}_{\text{n.c.}}$. There we use the infinite density matrix renormalization group (iDMRG) to approximate the ground state of a quantum spin chain whose long-wavelength behavior is described by TLL theory, weakly measure this state, and find that the behavior of correlation functions agrees with that predicted above.

In this section we have so far shown that, for $K<1$, arbitrarily weak measurements conspire at long wavelengths to suppress density fluctuations and to enhance phase fluctuations. For $K>1$ it is perhaps more striking to consider the regime of strong (but not quite projective) density measurements, corresponding to large but finite $v$. The implications of Eqs.~\eqref{eq:flowu} and \eqref{eq:flowy} are that these measurements fail to alter the long-distance behavior of correlation functions relative to the TLL ground state. This analysis has served to demonstrate the ideas and some of the techniques that can be used to study the effects of measurements on many-body quantum states. An important simplification was to restrict to a translation-invariant set of measurement outcomes; this allowed us to establish a connection to the behavior of an equilibrium system with a defect. Since the connection comes through an exchange of the roles of space and imaginary time, translation invariance in our problem corresponds to a time-independent coupling to the defect. Before proceeding to discuss the full ensemble of post-measurement states, let us comment briefly on how $\ket{\psi_{\text{n.c.}}}$ might be characterized in experiment.

\subsection{Coarse-grained correlations}\label{sec:coarse}

Throughout this section we have studied a quantum state $\ket{\psi_{\text{n.c.}}}$ that is extraordinarily difficult to prepare: to obtain it we must repeat the experiment a number of times that is exponential in the system size. Let us suppose that we prepare it (or, as we discuss in Sec.~\ref{sec:individual}, something close to it) just once. Because the post-measurement state is translation invariant, we can still determine its correlations through a spatial average, and here we study this average in detail. 

Recall that we expect $\braket{\nabla \hat \phi(x) \nabla \hat \phi(x')}_{\text{n.c.}}$ to decay as $|x-x'|^{-2}$ for $K>1$ and as $|x-x'|^{-2/K}$ for $K<1$. Unfortunately, the quantum-mechanical variance $V$ of this correlation function is dominated by short-wavelength fluctuations, and is therefore large compared to the expectation value. This fact prevents us from accurately determining $\braket{\nabla \hat \phi(x) \nabla \hat \phi(x')}_{\text{n.c.}}$ for a particular $x$ and $x'$ in a single run of the experiment.

Fortunately, because the operators $\nabla \hat \phi(x)\nabla \hat \phi(x')$ commute with one another for different values of $x$ and $x'$, we can measure all of them in a single run of the experiment. The spatial average of the correlation function suppresses the influence of quantum fluctuations. Let us define the spatially-averaged observable of interest as
\begin{align}
	\hat\Phi(x;W) = \frac{1}{W}\int_{0}^{W} dx' \nabla \hat \phi(x')\nabla \hat \phi(x'+x),
\label{eq:bigPhi}
\end{align}
where $W$ is the size of the region over which the signal is averaged. For $W$ large compared with the microscopic lengthscale the quantum variance of $\hat\Phi(x;W)$ is proportional to $V/W$. By contrast, its expectation value is independent of $W$. Therefore, for sufficiently large $W$ we can expect a measurement of the observable $\hat\Phi(x;W)$ in a single run to be representative of $\braket{\hat\Phi(x;W)}$. Note that for the transitions we have studied the required $W$ is polynomial in $x$, for example with $K<1$ where $\braket{\hat \Phi(x;W)} \sim -x^{-2/K}$, we require $W \gg V x^{4/K}$. For the phase correlations, an independent preparation of $\ket{\psi_{\text{n.c.}}}$ is required because $e^{i\hat \theta}$ and $\nabla \hat \phi$ do not commute. The operator $\cos[\hat \theta(x)-\hat \theta(x')]$ exhibits large quantum fluctuations in $\ket{\psi_{\text{n.c.}}}$, but we can again estimate its expectation value in a single run through a spatial average. In analogy with $\hat \Phi$ we define
\begin{align}
	\hat\Theta(x;W) = \frac{1}{W}\int_{0}^{W} dx' \cos[\hat \theta(x'+x)-\hat \theta(x')],
\label{eq:bigTheta}
\end{align}
and for large $W$ a single measurement of $\hat \Theta(x,W)$ is representative of its expectation value. Since $\braket{\hat \Theta(x;W)} \sim x^{-1/K}$ for $K<1$, the required $W \gg Vx^{2/K}$. For $K>1$ the algebraic decay of phase correlations $\braket{\hat \Theta(x;W)} \sim x^{-1/(2K)}$ is sufficiently slow that $\braket{\hat \Theta^2(x;W)} \sim W^{-1/(2K)}$, so in order to wash out the effects of quantum fluctuations we instead require ${W \gg x^2}$. 

Performing an average of a correlation function over space is natural when studying $\ket{\psi_{\text{n.c.}}}$ because it is translation-invariant. However, preparing such a state even once is exponentially costly. If we hope to observe a measurement-induced phenomenon without postselection, we must consider the structure of generic post-measurement states.

\section{Ensemble of outcomes}\label{sec:main}

In the previous section we discussed a transition in the structure of the state arising from one set of measurement outcomes. More generally, one is interested in the entire ensemble of measurement outcomes. In order to characterize fully the effect of our measurements, it is first necessary to identify physical quantities that encode the response of the quantum state. Second, we must average these over the ensemble of states arising from measurement. In this section we show that there is a transition, as a function of $K$, at the level of this ensemble. It is natural to expect that in this setting the critical Luttinger parameter is smaller than unity. This is because the postselection scheme in Sec.~\ref{sec:qnd} emphasises the role of density correlations relative to generic measurements. First, in Sec.~\ref{sec:nonlinear}, we introduce the measurement-averaged correlation functions of interest. Sections~\ref{sec:replicas} and \ref{sec:longwavelength} then describe a field-theoretic technique that allows us to analyze them. The transition is the subject of Sec.~\ref{sec:replicatransition}, and in Sec.~\ref{sec:individual} we discuss properties of individual postmeasurement states.

To follow this program it is useful to choose a different measurement model to Sec.~\ref{sec:qnd}; our choice will simplify the analytic calculation of ensemble-averaged correlation functions. First, we imagine coupling an observable $\hat m(x)$, which is a property of our measurement apparatus, to the normal-ordered density $\hat n(x)$. Second, we perform projective measurements of $\hat m(x)$ for all $x$. We denote by $m(x)$ the outcomes. This measurement protocol is implemented by an operator $\hat M_m$ that relates $\ket{\psi_{\text{g.s.}}}$ to the state $\ket{\psi_m}$ arising from the measurement outcomes $m(x)$,
\begin{align}
	\ket{\psi_m} &\equiv \braket{\hat M^2_m}^{-1/2}_{\text{g.s.}} \hat M_m \ket{\psi_{\text{g.s.}}},\notag\\
	\hat M_m &\propto e^{- \frac{1}{4} \mu \int dx [m(x) - \hat n(x)]^2},
\label{eq:Mm}
\end{align}
where $m(x)$ can take any real value, so here sums over outcomes $\sum_m$ should be interpreted as integrals. Note that the structure of $\hat M_m$ is strongly constrained by the requirement that the set of all $\hat M_m$ constitutes a quantum channel, $\sum_m \, \hat M^2_m=1$ (this condition also determines the prefactor in the second line of Eq.~\eqref{eq:Mm}), with the probabilities of different outcomes set by $p_m = \braket{\hat M^2_m}_{\text{g.s.}}$. We refer to this as the Gaussian measurement scheme, and discuss an implementation of $\hat M_m$ in Eq.~\eqref{eq:Mm} using ancillary quantum harmonic oscillators (QHOs) in Appendix~\ref{sec:gaussian}. To demonstrate the behavior of $\hat M_m$, consider first the case of an initial state $\ket{n}$ with definite densities: $\hat n(x)\ket{n} = n(x)\ket{n}$. We would then find from $p_m = \braket{n|\hat M^2_m |n}$ that the measurement outcomes $m(x)$ are normally-distributed around $n(x)$ with variance $\mu^{-1}$. In the following we will refer to the parameter $\mu$ as the measurement strength, with $\mu \to \infty$ the projective limit. For small $\mu$, the outcomes $m(x)$ are weakly correlated with $\braket{\hat n(x)}_m \equiv \braket{\psi_m|\hat n(x)|\psi_m}$. 

\subsection{Nonlinear observables}\label{sec:nonlinear}

To quantify the response of $\ket{\psi_{\text{g.s.}}}$ to local measurements, we average density and phase correlations over the ensemble of $\ket{\psi_m}$, and weight the results by the Born probabilities $p_m$. Because we perform this average, correlation functions that are sensitive to the nonlocal effects of measurements must be nonlinear in $\ket{\psi_m}\bra{\psi_m}$. To see why, consider for example
\begin{align}
	\braket{\hat n(0) \hat n(x)}_m \equiv \braket{\psi_m |\hat n(0) \hat n(x)|\psi_m}.
\end{align}
If we average $\braket{\hat n(0) \hat n(x)}_m$ over $m$, we find
\begin{align}
	\sum_m \, p_m \braket{\hat n(0) \hat n(x)}_m = \braket{\hat n(0) \hat n(x)}_{\text{g.s.}},
\label{eq:linearaverage}
\end{align}
which follows from $[\hat M_m,\hat n(0) \hat n(x)]=0$ and ${\sum_m \, \hat M^2_m=1}$. That is, the average of $\braket{\hat n(0) \hat n(x)}_m$ over measurement outcomes is totally insensitive to the fact that we measured the system. On the other hand, $\braket{\hat n(0) \hat n(x)}_m \neq \braket{\hat n(0) \hat n(x)}_{\text{g.s.}}$ in general, so the average in Eq.~\eqref{eq:linearaverage} has failed to capture the effect of our measurements. Another route to this fact comes from observing that averaging expectation values over the outcomes of density measurements is equivalent to dephasing in the basis of density eigenstates, and this does not affect density correlations. 

More generally, the averaged behavior of any correlation function linear in the density matrix cannot change in response to local measurements which act on different sites to the operators in the correlation function. The key point is that averaging over a measurement outcome is equivalent to replacing the measurement with a local quantum channel, which could just have well been implemented using a local unitary and an ancillary degree of freedom. The effects of local unitary operations are strictly local, and therefore so are the effects of measurements on averages of correlation function that are linear in post-measurement density matrices.

The correlation functions that we focus on are those of $\nabla \hat \phi$ and $e^{i\hat \theta}$. In particular, we probe correlations between quantum fluctuations of the density via
\begin{align}
	C(x) \equiv \sum_m \, p_m &\Big\langle \big[ \nabla \hat \phi(0) - \braket{\nabla \hat \phi(0)}_m \big] \label{eq:defC}\\ &\times\big[\nabla \hat \phi(x) - \braket{\nabla \hat \phi(x)}_m\big] \Big\rangle, \notag
\end{align}
and the phase through 
\begin{align}
	D(x) \equiv \sum_m \, p_m \, \Big| \braket{e^{i[\hat \theta(x)-\hat \theta(0)]}}_m\Big|^2. \label{eq:defD}
\end{align}
The behavior of $C(x)$ provides information on correlations between quantum fluctuations of the smooth part of the density. For large $\mu$, $\ket{\psi_m}$ approaches an eigenstate of the density operators, so we expect $C(x) \to 0$ for $\mu \to \infty$. Since knowledge of the density is incompatible with knowledge of the phase, in the limit of large $\mu$ we similarly expect $D(x) \to 0$. However, these quantities are not straightforward to compute analytically. Because of this we use a replica trick, writing e.g. $D(x) = \lim_{N \to 1} D_N(x)$ with
\begin{align}
	D_N(x) = Z_N^{-1} \sum_m \, p^N_m \big| \braket{e^{i(\hat \theta(x)-\hat \theta(0))}}_m \big|^2 \label{eq:defDN},
\end{align} 
and analogously for $C_N(x)$, where the role of the partition function is played by
\begin{align}
Z_N = \sum_m \, p^N_m.
\end{align}
Because contributions from different measurement outcomes are in Eq.~\eqref{eq:defDN} weighted by $p^N_m$ as opposed to $p_m$, we can view these correlation functions for $N > 1$ as biasing the average toward the most likely outcomes.

In the following we will compute correlation functions of the form Eq.~\eqref{eq:defDN} by writing the expectation values $\braket{\ldots}_m$ and probabilities $p_m$ as path integrals over configurations of $N$ replica fields $\varphi_{\alpha}(x)=\phi_{\alpha}(x,\tau=0)$, with $\alpha = 0 \ldots (N-1)$, that each interact with the local measurement $m(x)$. Integrating out $m(x)$ generates an interaction that favors `locking' the replicas together at $\tau=0$. The strength of this coupling between replicas increases with increasing measurement strength $\mu$. Physically, relative variations of the fields $\varphi_{\alpha}$ encode quantum uncertainty in the ground state density, and the locking of these fields together for $\mu \neq 0$ corresponds to the suppression of this uncertainty due to measurement.

Note also that the `free energy' $F_N = (1-N)^{-1} \log Z_N$ has an information-theoretic interpretation as an entropy of the measurement outcomes. Taking the limit $\mu \to \infty$ as our reference, we have
\begin{align}
	F_N -F_{N,\infty} = (1-N)^{-1} \log \frac{\sum_m \, p^{N}_m}{\sum_m \, p_{m,\infty}^N},
\label{eq:freeenergy}
\end{align}
where $p_{m,\infty}$ denotes the distribution of measurement outcomes for $\mu \to \infty$. For general $N$ the quantity $F_N$ is a R\'enyi entropy, and in the replica limit $N \to 1$ it is the Shannon entropy. 

\subsection{Replica field theory}\label{sec:replicas}

Here we develop a replica field theory that can be used to calculate correlation functions such as $C_N(x)$, as well as the free energy $F_N$. First consider the structure of the probability density
\begin{align}
	p_m &= \braket{\psi|\hat M^2_m|\psi} = \frac{\text{Tr}[ e^{-\beta \hat H} \hat M^2_m ]}{\text{Tr}[ e^{-\beta \hat H}]},
\end{align}
where the $\beta \to \infty$ limit is implicit. As usual we write the projector onto the ground state $e^{-\beta \hat H}$ as an integral over the field $\phi(x,\tau)$ and integrate out fluctuations at $\tau \neq 0$. The result is
\begin{align}
	p_m &= \int D\varphi \, e^{-s_{\mu}[\varphi,m]}, \label{eq:p_smu} \\
	s_{\mu}[\varphi,m] &= s[\varphi] + \frac{1}{2}\mu \int dx [m(x) - n(x)]^2,\notag
\end{align}
with $s[\varphi]$ given in Eq.~\eqref{eq:smalls} and $n(x)$ as in Eq.~\eqref{eq:density} but with the operator $\hat \phi(x)$ appearing there replaced by the scalar field $\varphi(x)$. Here we have absorbed the constant $\text{Tr}[ e^{-\beta \hat H}]$ into the measure $D\varphi$. We can then write, for example,
\begin{align}
	Z_N = \sum_m \int \prod_{\alpha} D\varphi_{\alpha} e^{-\sum_{\alpha} s_{\mu}[\varphi_{\alpha},m]}, \label{eq:p_smuN}
\end{align}
where the replica fields $\varphi_{\alpha}$ appear, and the index $\alpha=0,\ldots,(N-1)$. At the level of Eq.~\eqref{eq:p_smuN} the fluctuations of the various $\varphi_{\alpha}$ are independent, but all interact with the same measurement field $m$. The choice of the Gaussian form for $\hat M_m$ in Eq.~\eqref{eq:Mm} now allows us to integrate over $m$. This yields
\begin{align}
	Z_N = \int \prod_{\alpha} D\varphi_{\alpha} e^{-s_N[\{\varphi_{\alpha}\}]},
\end{align}
where we have omitted an overall constant that does not affect expectation values. Here the action
\begin{align}
	s_N[\{\varphi_{\alpha}\}] &= \sum_{\alpha} s[\varphi_{\alpha}] \notag \\&+\frac{\mu}{2}\int dx \sum_{\alpha\beta} \big(\delta_{\alpha \beta} - N^{-1}\big)n_{\alpha} n_{\beta} \label{eq:sN} 
\end{align}
describes coupling between the $N$ replicas $\varphi_{\alpha}$. Note that fluctuations of the symmetric linear combination of fields $\sum_{\alpha} n_{\alpha}$ are not affected by measurement; after averaging over outcomes, the measurements only have the effect of locking fluctuations in the different replicas together.

Using the action Eq.~\eqref{eq:sN} we calculate $C_N(x)$ as the average of $\nabla \varphi_0(0) \nabla \varphi_0(x)-\nabla \varphi_0(0) \nabla \varphi_1(x)$ with respect to the statistical weight $e^{-s_N}$. The first term generates the average of $\langle \nabla \hat \phi(0) \nabla \hat \phi(x) \rangle_m$, while the second generates the average of $\langle \nabla \hat \phi(0) \rangle_m \langle \nabla \hat \phi(x) \rangle_m$. The choice of replicas $\alpha=0,1$ is of course arbitrary. We represent the average over $\varphi_{\alpha}$ configurations with doubled angular brackets,
\begin{align}
    \langle \langle \ldots \rangle \rangle_N = Z_N^{-1} \int \prod_{\alpha} D\varphi_{\alpha} [\ldots]  e^{-s_N[\{\varphi\}]},
\end{align}
so that
\begin{align}
	C_N(x) =  \langle \langle  \nabla \varphi_0(0) \nabla \varphi_0(x) - \nabla \varphi_0(0) \nabla \varphi_1(x) \rangle\rangle_N
\label{eq:AN2}
\end{align}
An additional step is required for the phase correlations $D_N(x)$ because $\hat M_m$ does not commute with the operator $\hat \theta(x)$, and we describe this in Appendix~\ref{sec:mcorrelations}. Next we will show that the second term in Eq.~\eqref{eq:sN} gives rise to a transition, occurring as a function of $K$, in correlation functions such as $C_N(x)$ and $D_N(x)$. Since the entropy of the measurement record $F_N$ is the logarithm of the generating function $Z_N$ for these correlation functions, it too is sensitive to the transition.

\subsection{Long wavelengths}\label{sec:longwavelength}

To facilitate the RG analysis we express the action (\ref{eq:sN}) in terms of the fields $\varphi_\alpha$. In particular we have 
\begin{align}
	n_{\alpha} n_{\beta} &= \frac{1}{\pi^2} \nabla \varphi_{\alpha} \nabla \varphi_{\beta} \label{eq:nalphanbeta}\\&+ \frac{1}{ 2 \pi^2} \cos [2 (\varphi_{\alpha}-\varphi_{\beta})] + \ldots, \notag
\end{align}
where the ellipses represent terms that vary with wavenumbers $2k_F$ and $4k_F$. The integration $\int dx \, n_{\alpha} n_{\beta}$ in Eq.~\eqref{eq:sN} will wash these out. Note then that the first term in Eq.~\eqref{eq:nalphanbeta} gives a contribution to the action of the form $\int dq \, q^2 \tilde \varphi_{\alpha}(q) \tilde \varphi_{\beta}(-q)$, and that this is irrelevant compared with the term $\int dq\, |q| |\tilde \varphi_{\alpha}(q)|^2$ in $s[\varphi_{\alpha}]$ that comes from the ground-state density fluctuations. 

This discussion implies that long-wavelength fluctuations of the fields $\varphi_{\alpha}$ are described by
\begin{align}
	s_N[\{\varphi_{\alpha}\}] &=  \sum_{\alpha} s[\varphi_{\alpha}] \label{eq:sN2} \\&- \frac{\mu}{2N \pi^2} \int dx \sum_{\alpha < \beta} \cos[2 (\varphi_{\alpha}-\varphi_{\beta})] + \ldots, \notag
\end{align}
where the ellipses represent irrelevant contributions. As required, only inter-replica fluctuations $\varphi_{\alpha}-\varphi_{\beta}$ are suppressed by measurement. From here on we are only concerned with the contributions to $s_N$ displayed in Eq.~\eqref{eq:sN2}.

The replica action Eq.~\eqref{eq:sN2} should be contrasted with Eq.~\eqref{eq:snc}, where there is just a single field $\varphi$ and a perturbation of the form $v\cos 2\varphi$. Under the RG, fluctuations of $\varphi$ there have the effect of suppressing $v$ [see Eq.~\eqref{eq:flowu}]. In Eq.~\eqref{eq:sN2}, on the other hand, the perturbation $\mu \cos[2(\varphi_{\alpha}-\varphi_{\beta})]$ involves two fields $\varphi_{\alpha},\varphi_{\beta}$. Computing the scaling dimension of $\mu$ at first order in the perturbative RG, fluctuations of $\varphi_{\alpha}$ and $\varphi_{\beta}$ are independent. Consequently, the suppression of $\mu$ is twice as severe as that of $v$. Note that this is essentially the same computation as the one leading to Eq.~\eqref{eq:flowu}. The result is
\begin{align}
	\frac{d\mu}{d\ell} = (1-2K)\mu,
\label{eq:flowmu}
\end{align}
which reveals a fixed point at $K=1/2$, corresponding to strong repulsive interactions between fermions. For $K<1/2$ the effect of the measurements is relevant. Note also that, at first order in $\mu$, the critical Luttinger parameter $K=1/2$ for all integer $N \geq 2$.  As expected, the critical Luttinger parameter is here smaller than for the measurement scheme discussed in Sec.~\ref{sec:qnd}.

To discuss the opposite limit of large $\mu$ it is simplest to separate out the symmetric linear combination of fields, so we perform a Fourier transform over the replica index
\begin{align}
	\bar{\varphi}_{\kappa} = \sum_{\alpha} e^{i\kappa \alpha} \varphi_{\alpha}, \label{eq:replicaFourier}
\end{align}
where $\kappa= 0, 2\pi/N, \ldots, 2\pi(N-1)/N$ is an integer multiple of $2\pi/N$, and $\bar{\varphi}_{\kappa}(x)=\bar{\varphi}_{-\kappa}^*(x)$. In the case $N=2$ the action is 
\begin{align}
	s_{2}&[\{\bar{\varphi}_{\kappa}\}] =  \frac{1}{2}\sum_{\kappa} s[\bar{\varphi}_{\kappa}] - \frac{\mu}{4 \pi^2} \int dx \cos[2 \bar{\varphi}_{\pi}]. \label{eq:spm}
\end{align}
Clearly $\bar{\varphi}_0$ is unaffected by measurement. We can view the action for $\bar{\varphi}_{\pi}$ as having the same form as in Eq.~\eqref{eq:snc} although with a modified Luttinger parameter of $2K$. As in that case, the RG analysis of the strong-measurement limit of Eq.~\eqref{eq:spm} recovers the same critical $K$ as for weak measurements. Therefore, in the $N=2$ replica theory we find a critical $K=1/2$ for both weak and strong measurements. The analysis of the large-$\mu$ limit for $N > 2$ differs from that of the large-$v$ limit in Sec.~\ref{sec:qnd}, and we defer a detailed consideration of this regime to future work. The RG flow of $\mu$ for $N=2$ is shown in Fig.~\ref{fig:RGreplica}(a).

\subsection{Transition}\label{sec:replicatransition}

\begin{figure}
\includegraphics[width=0.47\textwidth]{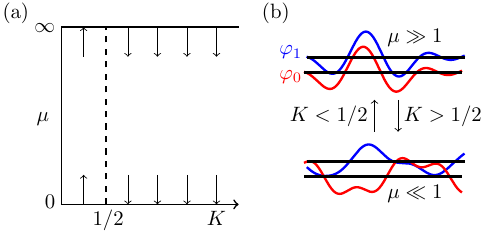}   
\caption{Transition in the theory described by $s_N$ as a function of $K$. (a) RG flow of measurement strength $\mu$. For small $\mu$ the linearized RG flow is the same for all $N$, and changes direction at $K=1/2$. For large $\mu$ we have shown that the change in the direction of the flow is, for $N=2$, also at $K=1/2$. (b) Structure of the field configurations that dominate correlation functions in the case $N=2$. For $K<1/2$ and at long wavelengths $\varphi_0$ and $\varphi_1$ are locked together by $\mu \cos[2(\varphi_0-\varphi_1)]$, whereas for $K>1/2$ their fluctuations are independent. As in Fig.~\ref{fig:RGnc} the directions of arrows on the right indicate the effect of coarse-graining.}
\label{fig:RGreplica}
\end{figure}

Here we discuss the nature of the transition at $K=1/2$ by comparing the structure of correlation functions in the two phases. First, note that the replica action $s_N[\varphi_0 \ldots \varphi_{N-1}]$ in Eq.~\eqref{eq:sN} is invariant under the exchange of fields $\varphi_{\alpha} \leftrightarrow \varphi_{\beta}$. However, for $K<1/2$ where $\cos[2(\varphi_{\alpha}-\varphi_{\beta})]$ is relevant, we expect the path integral to be dominated by field configurations with $\varphi_{\alpha}-\varphi_{\beta}$ an integer multiple of $\pi$. This suggests a spontaneous breaking of the exchange symmetry at small $K$. Each of the symmetry-broken configurations can be labeled by a set of $N-1$ integers $p_{\alpha} = \pi^{-1}(\varphi_{0}-\varphi_{\alpha})$ with $\alpha \geq 1$. To understand them, it is helpful to consider a domain wall, for example a sharp increase in $p_{\alpha}$ around $x=0$. Jumps in $\varphi_{\alpha}$ by $\pi$ do not alter the local density away from the jump. Instead, an increase of $\varphi_{\alpha}$ by $\pi$ corresponds to a missing particle in the vicinity of the jump. This means that a domain wall across which $p_{\alpha}-p_{\beta}$ increases by an integer corresponds to a decrease in the local particle density in replica $\alpha$ relative to replica $\beta$. For $K<1/2$, under coarse-graining these domain walls become dilute, indicating that long-wavelength features in the particle density match across the different replicas. For $K>1/2$, fluctuations of the fields $\varphi_{\alpha}$ are independent in the limit of long wavelengths. We illustrate these two different behaviors in Fig.~\ref{fig:RGreplica}(b).

The correlation functions $C_N(x)$ and $D_N(x)$ describe fluctuations between the different replicas. For $N=2$ they can be expressed as
\begin{align}
	C_2(x) &= \frac{1}{2}\langle \langle \nabla \bar{\varphi}_{\pi}(0) \nabla \bar{\varphi}_{\pi}(x) \rangle \rangle_2 \\
	D_2(x) &= \langle \langle e^{i[\bar{\theta}_{\pi}(x)-\bar{\theta}_{\pi}(0)]} \rangle \rangle_2, \notag
\end{align}
where $\bar{\theta}_{\pi}=\theta_0-\theta_1$ describes the phase difference between the replicas. These correlation functions can be calculated by analogy with Eqs.~\eqref{eq:ncgradphi} and \eqref{eq:nctheta}, respectively, and the fields $\bar{\varphi}_{\pi}$ and $\bar{\theta}_{\pi}$ can be viewed as experiencing effective Luttinger parameters $2K$ and $K/2$. For $K>1/2$ we then have
\begin{align}
	C_N(x) \sim -x^{-2}, \quad D_N(x) \sim x^{-1/K},
\end{align}
which is the same behavior as in $\ket{\psi_{\text{g.s.}}}$. For $K<1/2$ the discussion in Sec.~\ref{sec:qnd} implies the behavior of $C_2(x)$ and $D_2(x)$; after changing variables to $\bar{\varphi}_{0}$ and $\bar{\varphi}_{\pi}$, it is clear that the saddle points of $s_2$ have essentially the same structure as those of $s_{\text{n.c.}}$. These saddle points correspond simply to configurations of domain walls in the $\bar{\varphi}_{\pi}$ field, and the leading contribution at large $x$ comes from a pair of oppositely-signed walls at separation $x$. The results are
\begin{align}
	C_2(x) \sim -x^{-1/K}, \quad D_2(x) \sim x^{-2/K},
\end{align}
with prefactors set by the square of the domain wall fugacity. For large $\mu$ no rescaling is necessary and this fugacity is exponentially small in $\mu^{1/2}$ [see Appendix~\ref{sec:walls}]. This is consistent with our expectation that each of $C(x)$ and $D(x)$ should vanish in the limit of projective measurements $\mu \to \infty$. For $N>2$, although the saddle points of $s_N$ have more structure, it is natural to expect that for large $\mu$ they also have an interpretation as domain walls in real space. In the replica theory these are domain walls separating different ways of locking the fields $\varphi_{\alpha}$ to one another. 

To develop a physical interpretation for the faster decay of $C_2(x)$ for $K<1/2$, let us recall the definition of $C(x)$ in Eq.~\eqref{eq:defC}. This correlation function can be expressed as the measurement-averaged difference between the two-point function $\braket{\nabla \hat\phi(0)\nabla \hat\phi(x)}_m$ and the product of one-point functions $\braket{\nabla \hat\phi(0)}_m\braket{\nabla \hat\phi(x)}_m$ and, in a disentangled eigenstate of the density operators, this difference must vanish. We can therefore understand the faster decay of $C(x)$, occurring when density measurements are relevant, as capturing the approach to behavior resembling that in a product state. 

In this section we have investigated the difference between $K<1/2$ and $K>1/2$, where generic measurement outcomes respectively behave as relevant and irrelevant perturbations. For all values of $K$ we expect that correlations of the density and phase remain algebraic, although the exponents characterizing these decays change their dependence on $K$ across the transition. For the correlation functions that we have considered, the decay is more rapid for $K<1/2$. In this case arbitrarily weak local measurements are sufficient to alter the structure of the quantum state at the longest wavelengths and distances. For $K>1/2$, as long as $\mu$ is finite, the long distance behavior of the correlations is unchanged compared to the unmeasured ground state. Note, however, that the results obtained above for correlation functions in the regime $K<1/2$ are strictly appropriate only for $N=2$. This corresponds to averaging nonlinear correlation functions over the ensemble of measurement outcomes with $p^2_m$ weights, as opposed to the Born probabilities $p_m$ relevant for experiment. 

\subsection{Individual outcomes}\label{sec:individual}
While there is analytic simplicity only in averages over the ensemble of measurement outcomes, it is natural to ask whether we can say anything concrete about the structure of individual post-measurement states. Here we will briefly consider the problem of evaluating expectation values in a typical state $\ket{\psi_m}$. As an example we have
\begin{align}
	\braket{\hat n(0)}_m = \frac{\int D\varphi \, n(x) \, e^{-s_{\mu}[\varphi,m]}}{\int D\varphi \, e^{-s_{\mu}[\varphi,m]}}.
 \label{eq:individual1}
\end{align}
From this expression we see that, in order to develop some intuition for the post-measurement state $\ket{\psi_m}$, we only need consider the action $s[\varphi] - \frac{1}{2}\mu\int dx\, m(x)n(x)$, where $m(x)$ appears as a field coupling to the density $n(x)$. This is because the contribution to Eq.~\eqref{eq:p_smu} proportional to $\int dx\, m^2(x)$ cancels between numerator and denominator, while the (ill-defined) contribution proportional to $\int dx\, n^2(x)$ can be removed by starting from a concrete lattice model and subsequently taking the continuum limit [see Appendix~\ref{sec:gaussian}]. 

When discussing the coupling between $m(x)$ and $n(x)$ it is important to note that, although the measurement outcomes $m(x)$ are to some extent random, they have power-law correlations inherited from the ground state. Explicitly,
\begin{align}
    \sum_m p_m m(0)m(x) = \braket{\hat n(0) \hat n(x)}_{\text{g.s.}} + \mu^{-1}\delta(x),
\label{eq:mmcorrelations}
\end{align}
 which can be shown directly from Eq.~\eqref{eq:Mm}. Because the correlations of $n(x)$ are most easily understood through the decomposition into smooth and oscillatory components in Eq.~\eqref{eq:density}, we will do the same for $m(x)$, i.e. 
\begin{align}
    m(x) = m_0(x) + [m_{2k_F}(x)e^{2ik_F x} + \text{c.c.}],
\label{eq:mdecompose}
\end{align}
where $m_{2k_F}(x)$ is in general complex. The two-point functions $\sum_m p_m m_0(0)m_0(x)$ and $\sum_m p_m m_{2k_F}(0)m_{2k_F}(x)$ can now be understood as reproducing the ground state correlations $\braket{\nabla \hat \phi(0) \nabla \hat \phi(x)}_{\text{g.s.}}$ and $\braket{\cos [2 \hat \phi(0)] \cos [2\hat \phi(x)]}_{\text{g.s.}}$, respectively. Inserting Eqs.~\eqref{eq:density} and \eqref{eq:mdecompose} into the perturbation arising from measurements, and choosing a field $m_{2k_F}(x)$ that is real for simplicity, we find 
\begin{align}
    \int dx \, m(x) n(x) = \int dx \, m_{2k_F}(x) \cos[2\varphi(x)] + \ldots,
\end{align}
where on the right-hand side we have chosen to display only the term that controls the transition studied in this section. The others can be seen to give rise to irrelevant perturbations.

To assess the effect of a typical measurement outcome $m(x)$ we adapt the real-space RG argument in Ref.~\cite{andelman1984scale} to power-law correlated fields. Recall that the scaling dimension of $\cos[2\varphi(x)]$ at the $\mu=0$ fixed point is $K$ so that, under a change of the microscopic length scale by $b$, a uniform $m_{2k_F}(x)$ would be renormalized by a factor $b^{1-K}$ [see Eq.~\eqref{eq:flowu}]. More generally, we must first ask how the magnitude of the coarse-grained field $\frac{1}{b}\int_{x-b/2}^{x+b/2}dx'\,m_{2k_F}(x')$ varies with $b$. The average of this field over $m_{2k_F}(x)$ is zero, so we must instead evaluate the average of its square. Using the fact that correlations of $m_{2k_F}(x)$ are inherited from those of $\cos[2\varphi(x)]$ [see Eq.~\eqref{eq:mmcorrelations}] we find
\begin{align}
    \sum_m p_m \Big|\frac{1}{b}\int_{x-b/2}^{x+b/2} dx' m_{2k_F}(x')\Big|^2 \sim b^{-2K}, \label{eq:averagem2}
\end{align}
for $K<1/2$, since then the double integral over positions is dominated by points separated by $\sim b$. For $K>1/2$ this integral is instead dominated by small separations and we recover the result for uncorrelated random fields $b^{-1}$.

Equation~\eqref{eq:averagem2} shows that coarse-graining $m_{2k_F}(x)$ over a length scale $b$ typically suppresses the amplitude of its fluctuations by a factor $b^{-K}$. The coarse-grained $m_{2k_F}(x)$ within the interval $[x-b/2,x+b/2]$ can then be viewed as a uniform field coupled to $b^{-1}\int_{x-b/2}^{x+b/2} dx\,\cos[2\varphi(x')]$, which is simply the coarse-grained $\cos[2\varphi(x)]$. Therefore, when we eliminate fluctuations on length scales smaller than $b$, we should complete the RG transformation by rescaling the averaged field as if it were uniform \cite{andelman1984scale}, i.e. by a factor $b^{1-K}$. From this we find that, for $K<1/2$, the overall rescaling of the measurement strength is $\mu \to b^{1-2K}\mu$, consistent with result Eq.~\eqref{eq:flowmu} obtained from our replica approach. Note that for $K>1/2$, where averaging $m_{2k_F}(x)$ suppresses the amplitude of its fluctuations by a factor $b^{-1/2}$ as for random fields, the above line of reasoning leads instead to $\mu \to b^{\frac{1}{2}(1-2K)}\mu$. 

In closing this section we note that the above reasoning applies to general critical ground states. For a critical state in $d$ spatial dimensions, where we can view ground-state correlations as $\tau=0$ correlations in a $(d+1)$-dimensional field theory, measurements appear as a power-law correlated random field on the $\tau=0$ surface. Correlations in an individual post-measurement state then correspond to $\tau=0$ correlations in the presence of this surface field. Generalizing the above real-space RG arguments to $d$ spatial dimensions, and to the measurement of an operator with a scaling dimension which we now denote $\Delta$ to avoid confusion, for $\Delta<d/2$ the measurement strength is rescaled as $\mu \to b^{d-2\Delta}\mu$. For $\Delta>d/2$ it is instead rescaled as $\mu \to b^{\frac{1}{2}(d-2\Delta)} \mu$. This argument suggests that, if uncorrelated random fields on a surface of codimension one are relevant, so are measurements.

Finally we comment on the relation between the results of this section and of those in Sec.~\ref{sec:qnd}, where we considered the effect of postselecting for a particular set of outcomes. In the language of this section, the theory in Sec.~\ref{sec:qnd} corresponds to uniform $m_{2k_F}(x)$, and this corresponds to a relevant perturbation for $K<1$. It is natural to ask whether such extreme postselection is necessary to generate a perturbation of this kind, and the answer is in the negative. This is because randomness in the field $m_{2k_F}(x)$ is irrelevant for all $K>1/2$, and so will not affect a transition occurring at $K=1$. In the Gaussian scheme, for a field $m_{2k_F}(x)$ whose average over a spatial region is independent of the size of that region, the quantum state should therefore be restructured in the manner described in Sec.~\ref{sec:qnd}. Based on this it is natural to conjecture that, within the measurement scheme with binary outcomes discussed in that section, the transition should be robust to a small finite density of `clicks'.

\section{Avoiding postselection}\label{sec:postselection}
A barrier to experimental studies of the effects of measurements on quantum systems is that signatures are only to be found in physical quantities conditioned on the measurement outcomes. In this section we discuss this `postselection problem' \cite{skinner2019measurement,li2019measurement,gullans2020scalable}. To understand the origin of the postselection problem, let us consider the scenario where a quantum state $\ket{\psi_m}$ is prepared by a sequence of $M$ measurements. Given $\ket{\psi_m}$ we then try to characterize its structure by estimating the expectation value of a `probe' observable. To do this, we have to measure the probe observable, and in a given run of the experiment we can only obtain one result (i.e., one of its eigenvalues). Determining the expectation value of the probe observable in $\ket{\psi_m}$ requires us to repeat the experiment, but this is very resource intensive: the probability that we successfully prepare $\ket{\psi_m}$ again is exponentially small in $M$, and so for large $M$ we are unlikely to ever prepare $\ket{\psi_m}$ again. In this section we show that the effects of measurements can nevertheless be observed when one has access to an appropriate simulation on a classical computer.

As an example of the problem at hand, let us first suppose that our aim is to determine the nonlinear contribution to $C(x)$ in Eq.~\eqref{eq:defC}, which is  $\sum_{m} p_m \braket{\nabla \hat \phi(0)}_{m} \braket{\nabla \hat \phi(x)}_{m}$. In a given run of the experiment in which we find outcomes $m$, we can try to estimate the post-measurement expectation values of the probe observables $\nabla \hat \phi(0)$ and $\nabla \hat \phi(x)$, and to do this we must measure them. The results are eigenvalues of $\nabla \hat \phi(0)$ and $\nabla \hat \phi(x)$, and so our best estimate for $\braket{\nabla \hat \phi(0)}_{m} \braket{\nabla \hat \phi(x)}_{m}$ is the product of these eigenvalues. However, this product of eigenvalues is also an eigenvalue of $\nabla \hat \phi(0) \nabla \hat \phi(x)$. Performing an average over many runs of the experiment, we therefore find convergence to $\sum_{m} p_m \braket{\nabla \hat \phi(0)\nabla \hat \phi(x)}_{m}$ instead of the desired quantity $\sum_{m} p_m \braket{\nabla \hat \phi(0)}_{m} \braket{\nabla \hat \phi(x)}_{m}$. The result is therefore an average of a quantity that is linear in the post-measurement density matrix, and corresponds to the expectation value of $\nabla \hat \phi(0)\nabla \hat \phi(x)$ in the case where the state is first dephased in the measurement basis. This example illustrates the fact that, if we do not use the information obtained from measurement, simple averages over outcomes do not distinguish measurement from dephasing. 

In the simplest scenario the data available consists of (i) the outcomes $m$, assumed different for each run of the experiment, and (ii) one eigenvalue $\lambda_m$ of a probe observable for each of these outcomes. Given this data, we must ask more generally which kinds of physical quantities we can determine, and one possibility is to compute the average of $w_m\lambda_m$ over runs of the experiment, where $w_m$ is an $m$-dependent weight that we are free to choose (note that if $w_m=1$ our average reproduces the effects of dephasing). This average over experimental runs washes out quantum fluctuations, and so the result of this protocol converges to e.g. $\sum_m p_m w_m \braket{\nabla \hat \phi(x)}_m$ in the case where $\nabla \hat \phi(x)$ is the probe observable. 

For concreteness let us consider again the nonlinear contribution to $C(x)$ in Eq.~\eqref{eq:defC}, a product of post-measurement expectation values $\braket{\nabla \hat \phi(0)}_{m}$ and $\braket{\nabla \hat \phi(x)}_{m}$ averaged over runs of the experiment. Now suppose that these expectation values can be estimated from a calculation on a classical computer; we denote these estimates by e.g. $\braket{\nabla \hat \phi(0)}_{m,C}$, and they are to be distinguished from the true `quantum' expectation values $\braket{\nabla \hat \phi(0)}_m$. We now propose that a physically-meaningful choice for the weighting $w_m$ is the classically-estimated expectation value $w_m = \braket{\nabla \hat \phi(0)}_{m,C}$.

This leads us to define `quantum-classical' estimators as averages of such $w_m \lambda_m$ over runs of the experiment. For example, if we want to construct the quantum-classical estimator for the nonlinear contribution to $C(x)$, we can choose $\nabla \hat \phi(x)$ as our probe observable and $w_m = \braket{\nabla \hat \phi(0)}_{m,C}$ as our weighting; this quantum-classical estimator converges to
\begin{align}
    \sum_m p_m  \braket{\nabla \hat \phi(0)}_{m,C} \braket{\nabla \hat \phi(x)}_{m}\label{eq:QC}
\end{align}
and describes cross-correlations between experiment and the classical calculation. This is an unusual situation where, although it is not possible to directly compare experiment with simulation for large $M$, one can compare their cross-correlation with the simulation, i.e. one can compare the quantum-classical estimator in Eq.~\eqref{eq:QC} with the `classical-classical' estimator $\sum_m p_m  \braket{\nabla \hat \phi(0)}_{m,C} \braket{\nabla \hat \phi(x)}_{m,C}$.

If the classical-classical probe changes its behavior as a parameter (such as $K$) is tuned, and if it coincides with the quantum-classical probe, this provides evidence for a restructuring of the experimentally-prepared quantum state. Crucially, this is a signature that does not suffer from a postselection problem. We note also that, for the particular problems that we have discussed in this work, the unmeasured quantum state can be well-approximated by a matrix product state (MPS) having bond dimension $\chi$ that is polynomial in the system size $L$ \cite{verstraete2006matrix}. Therefore, the classical memory requirements $\sim L\chi^2$ for constructing quantities such as the one displayed in Eq.~\eqref{eq:QC} are themselves polynomial in $L$.

The idea of using classical simulations to construct probes of the effects of measurement on many-body states was previously used in Ref.~\cite{gullans2020scalable,dehghani2022neural} in the context of the dynamical MIPT, although the way the classical information is processed is in that approach quite different. The quantum-classical estimators above do however have an interesting parallel in the cross entropy used to demonstrate quantum supremacy in Ref.~\cite{arute2019quantum}. A key distinction is that here we are advocating for their use as observables in their own right, rather than as benchmarks for a quantum simulation.

The quantum-classical estimator will of course be noisier than the classical-classical one, simply because the former is affected by quantum fluctuations while the latter is not. A more concerning source of error is mismatch between the real quantum system and the classical representation of its state. To reduce these differences, one possibility in the case where the initial state is translation invariant is to coarse-grain the observables of interest. In addition to reducing quantum fluctuations, it is natural to expect that an averaging procedure of this kind will suppress the effects of microscopic differences between the quantum system and the classical approximation. The limitations of this approach will depend sensitively on the experimental system of interest, and we defer a detailed investigation to future work.

\section{Discussion}\label{sec:discussion}

Questions about the effects of observation on many-body quantum states become ever more pertinent as quantum computation and simulation technologies develop. Critical states are of particular interest in this context since they are highly entangled. In this work we have shown that local measurements performed on critical quantum ground states can conspire with one another to drive transitions in long-wavelength correlations. Such an instantaneous restructuring of the quantum state is possible due to the nonlocality of the measurement process, and the algebraic correlations characteristic of critical states. 

A central result is to demonstrate how the nonlocal effects of measurements can be understood using standard tools from quantum statistical mechanics. In this language, measurement-induced transitions in ground state correlations map to boundary phase transitions. The bulk corresponds to the Euclidean action generating ground-state correlations, while the measurements are a boundary perturbation appearing at a fixed imaginary time. This interpretation is quite general, and in higher spatial dimensions it implies a relation between surface critical phenomena and transitions in the structure of quantum states.

Our focus here has been on one-dimensional quantum liquids, in particular spinless TLLs. We have shown that transitions in the structure of the weakly-measured ground state occur as the Luttinger parameter $K$ is varied. First, in Sec.~\ref{sec:qnd}, we mapped the calculation of correlation functions in a particular measured state onto the problem of a local potential defect in a TLL \cite{kane1992transport,kane1992transmission}. The state we chose was translation invariant, and in the dual defect problem this property corresponds to a potential barrier that does not vary in (imaginary) time. For $K<1$ we showed that for arbitrarily weak coupling between the quantum system and the measurement apparatus there is a change in the form of correlation functions at long distances. In particular, density fluctuations and phase correlations are suppressed. These effects are manifest in faster power-law decays of $\nabla \hat \phi$ and $e^{i\hat \theta}$ correlations. For $K>1$, the measurements are irrelevant, in the sense that they do not alter the correlations on large scales.

Following this, we investigated the full ensemble of quantum states that can arise from measurement. To make analytic progress, we averaged physical quantities over this ensemble. In order to distinguish the effects of measurement from the effects of coupling to an environment, it is necessary for these quantities to be nonlinear in the system density matrix. To calculate their averages we developed a replica field theory, within which measurements act as a coupling between the different replicas in all space but only at a single imaginary time $\tau=0$. In this formulation the question is whether the coupling is relevant. For the density measurements that we considered, we found that it is relevant for $K<1/2$ and irrelevant for $K>1/2$. As $K$ is decreased, the measurements drive a transition which breaks the symmetry of the theory under the exchange of replicas, and which has signatures in the asymptotic forms of averaged nonlinear correlation functions. For $K>1/2$ density correlations in the initial quantum state are too weak, and measurements fail to restructure it.

Within our replica framework, questions remain over the behavior of the theory in Eq.~\eqref{eq:sN2} for $K < 1/2$ and $N \neq 2$. To answer these, one must presumably account for the structure of the saddle points at large $\mu$. Doing so may allow for the calculation of the averaged nonlinear correlation functions \eqref{eq:defC} and \eqref{eq:defD} in the replica limit. Replicas are, however, just one of a number of possibilities when studying an ensemble of random outcomes. Another is to adapt supersymmetric methods from the study of disordered systems~\cite{efetov1983supersymmetry}, although these are unlikely to be appropriate in our problem since the density operator $\hat{n}$ is nonlinear in $\hat{\phi}$. A third possibility is to approach the problem numerically. In Appendix~\ref{sec:numerics} we have used iDMRG to calculate correlation functions in the translation-invariant state $\ket{\psi_{\text{n.c.}}}$, but tensor-network techniques also open the door to the study of nonlinear correlation functions in generic measured quantum states, and to averages weighted with respect to the Born probabilities $p_m$ (as opposed to $p^N_m$). One could otherwise tackle these problems using quantum Monte Carlo \cite{sandvik2010computational}, here applicable since we are only concerned with imaginary-time evolution.

An important question is whether the phenomena we have studied can be observed in experiment. In discussing this, it is useful to recall the barriers to observations of dynamical MIPTs. One is the necessity to postselect on individual measurement trajectories \cite{skinner2019measurement,li2019measurement,gullans2020scalable,ippoliti2021postselection}. This problem arises because, when characterizing a quantum state prepared by measurements, the experimenter ultimately has to measure an observable, and this process is destructive. In one run of an experiment, a given observable can only be measured once, but if the quantum fluctuations of the observable are large the result of this measurement is a poor estimate for its expectation value. To estimate the latter, the same state has to be prepared a number of times, but the probability for its successful preparation is in general exponentially small in the number of measurements required to do so. Since the dynamical MIPT occurs in the limit of large times $t$ and system sizes $L$ with $t \propto L$, the number of measurements required scales as $L^2$, and hence the number of experimental runs required is astronomical even for moderate $L$. The postselection requirements are in our case less severe, since the number of measurements is of order $L$ rather $L^2$, although an exponential-in-$L$ postselection overhead is still prohibitive. As we have discussed in Sec.~\ref{sec:coarse} the possibility for spatial averaging does allow for quantum fluctuations to be suppressed in individual runs of an experiment, but this does not remove the exponential overhead.

However, if it is possible to determine conditional expectation values of observables classically, we have shown in Sec.~\ref{sec:postselection} that cross-correlating the results of these calculations with experimental data alleviates the postselection problem. This idea has parallels in the context of the dynamical MIPT in stabilizer circuits, where one can avoid postselection by determining a `decoder function' \cite{gullans2020scalable,dehghani2022neural}. In this work we have been concerned with critical states in one spatial dimension, so the computational resources required are only polynomial in the system size. Because of this, cross-correlations can in principle be constructed for experiments performed on quantum simulators using hundreds if not thousands of qubits.

In addition to this dramatic reduction in experimental resource requirements, we note that measurement-induced phenomena in this static setting are far less sensitive to decoherence than those arising in dynamical systems. This is because local quantum channels have strictly local effects in the absence of any subsequent dynamics. In particular, correlations between local observables are not affected by channels acting elsewhere in the system, but they are nonlocally affected by measurements. The results of this work therefore represent a significant advance toward the observation of measurement-induced phenomena. The already rich history of experiments on TLL behavior in ultracold atomic gases \cite{paredes2004tonks,kinoshita2004observation,hofferberth2008probing,yang2017quantum}, as well as developments in quantum-gas microscopy \cite{bakr2009quantum,gross2021quantum}, makes this class of systems a promising physical setting. 

\acknowledgements
We thank Yimu Bao, Michael Buchhold, John Chalker, Yaodong Li, David Luitz, Adam Nahum, Chandra Varma and Yantao Wu for useful discussions, as well as Michael Gullans and Matthew Fisher for valuable comments on the manuscript. This work was supported by the Gordon and Betty Moore Foundation (SJG), UC Berkeley Connect (ZW), the Gyorgy Chair in Physics at UC Berkeley (EA), and in part by the NSF QLCI program through grant number OMA-2016245.

\newpage
\appendix

\onecolumngrid

\section{Domain walls}\label{sec:walls}
Here we discuss the description of the large-$v$ limit of the theory Eq.~\eqref{eq:snc} in terms of domain walls. When $v$ is large we typically have $\varphi \simeq p\pi$ for integer $p$, and the integer jumps of $p$ are domain walls. Formally, this is a description of the saddle-point approximation to the partition function $\int D\varphi e^{-s_{\text{n.c.}}[\varphi]}$, and the different saddle points corresponds to different domain-wall configurations. To describe domain walls it is necessary to first introduce a short-wavelength regularization, and we choose to add a term $\frac{1}{2}\int dx (\nabla \varphi)^2$. Writing $x = v^{-1/2} x'$ we find from Eq.~\eqref{eq:snc}
\begin{align}
	s_{\text{n.c.}}[\varphi] = s[\varphi] - v^{1/2} \Big[  \int dx' \cos 2\varphi - \frac{1}{2} \int dx' (\nabla' \varphi)^2 \Big]. \label{eq:sncrescaled}
\end{align}
For large $v$ we can neglect the contribution $s[\varphi]$ in the first instance. Varying the term in square brackets with respect to $\varphi$ we can then find the structure of the saddle points.  In the case of a single domain wall we set $\varphi(x') \to 0,\pi$ and $\nabla' \varphi(x') \to 0$ for $x' \to \mp \infty$, and the result is
\begin{align}
	\varphi_{\text{d.w.}}(x) = \frac{\pi}{2} + \tan^{-1}\big[ \sinh(2 v^{1/2} x)\big], \label{eq:saddlesolution}
\end{align}
which is a domain wall with width $\sim v^{-1/2}$. Inserting Eq.~\eqref{eq:saddlesolution} into Eq.~\eqref{eq:sncrescaled} and neglecting the contribution from $s[\varphi]$ we find $s_{\text{n.c.}}[\varphi_{\text{d.w.}}]-s_{\text{n.c.}}[\varphi=0] = 4v^{1/2}$, so $g \equiv \exp(-4v^{1/2})$ is the fugacity of a domain wall. 

For $K<1$ we will evaluate the asymptotic properties of correlation functions within the dilute domain wall approximation. To see why this description is possible, note first that if we were to neglect the interactions between domain walls then we would find that their typical separation is $\sim 1/g$. Comparing this with their width $\sim v^{-1/2}$ it is clear that for large $v$, and hence on large scales in the coarse-grained theory for $K<1$, we have $1/g \gg v^{-1/2}$. If we are interested only in correlations on scales much larger than $v^{-1/2}$, it suffices to approximate $\varphi$ as a sum of step functions $\varphi(x) = \pi \sum_j \epsilon_j \Theta(x-x_j)$, with $\epsilon_j = \pm 1$, as in Eq.~\eqref{eq:sharpwalls}. Inserting this expression into $s[\varphi]$ we find a long-range attractive interaction between domain walls with oppositely-signed $\epsilon_j$, and a long-range repulsion between those with the same sign. The partition function for the domain walls is then
\begin{align}
	Z_{\text{d.w.}} = \sum_{n=0}^{\infty} g^{2n} \int_{x_j < x_{j+1}} \prod_{j=1}^{2n} dx_j \sum_{\{\epsilon\}, \sum_j \epsilon_j=0} e^{\frac{2}{K} \sum_{j < k} \epsilon_j \epsilon_k \log|x_j-x_k|}. \label{eq:Zdw}
\end{align}
For simplicity we consider periodic boundary conditions in space, which gives the constraint $\sum_j \epsilon_j = 0$ on the sum over all possible $\epsilon_j = \pm 1$ configurations. This constraint implies that the number $2n$ of domain walls is even. To remain consistent with our approximation of dilute domain walls we should additionally restrict $|x_{j+1}-x_j| > v^{-1/2}$. 

From the theory Eq.~\eqref{eq:Zdw} we can determine the RG flow of the parameter $v$ in the regime of strong measurements, giving Eq.~\eqref{eq:flowy}. Although this is standard, we include it here for completeness. The key observation is that
\begin{align}
	Z_{\text{d.w.}} \simeq \int D\vartheta e^{- \frac{K}{4\pi} \int \frac{dq}{2\pi} |q| |\tilde \vartheta(q)|^2 + 2g \int dx \cos \vartheta(x)}, \label{eq:Zdwtheta}
\end{align}
where we have introduced a real scalar field $\vartheta(x)$. We demonstrate this connection below, but first note that the statistical weight in this expression has the same form as Eq.~\eqref{eq:snc} with the substitutions $\varphi \to \vartheta$, $K \to 4/K$, $v \to 2g$ and $\cos 2\varphi \to \cos\vartheta$. In the limit of small $g$, corresponding to large $v$, the scaling dimension of $\cos \vartheta$ can be determined in perturbation theory, and at first order we find $dg/d\ell = (1-1/K)g$. Using $g = \exp(-4v^{1/2})$ gives Eq.~\eqref{eq:flowy}. In the domain wall picture, for $K<1$ the attractive interaction between oppositely-signed domain walls is sufficiently strong that they become ever more dilute under coarse-graining. This manifests as a decrease in the fugacity. 

The connection between Eqs.~\eqref{eq:Zdw} and \eqref{eq:Zdwtheta} follows from an expansion of the latter in powers of $g$. Integrating this expansion over $\vartheta$ eliminates all terms featuring an odd number of cosines, leading to
\begin{align}
	Z_{\text{d.w.}} &\simeq \sum_{n=0}^{\infty} \frac{(2g)^{2n}}{(2n)!} \int \prod_{j=1}^{2n} dx_j \int D\vartheta \cos\vartheta(x_1) \ldots \cos\vartheta(x_{2n}) e^{- \frac{K}{4\pi} \int \frac{dq}{2\pi} |q| |\tilde \vartheta(q)|^2} \notag \\
	&= \sum_{n=0}^{\infty} g^{2n} \int_{x_j < x_{j+1}} \prod_{j=1}^{2n} dx_j \sum_{\{\epsilon\}, \sum_j \epsilon_j=0} e^{\frac{2\pi}{K} \sum_{j<k} \epsilon_j \epsilon_k \int \frac{dq}{2\pi}\frac{1}{|q|} (1-\cos[q(x_j-x_k)])}, 
\end{align}
where in the second line we have ordered the sum so that $x_j < x_{j+1}$, thereby cancelling the factor $(2n)!$. We have also used $\cos\vartheta(x_j) = \frac{1}{2}\sum_{\epsilon_j=\pm 1} e^{i\epsilon_j \vartheta(x_j)}$, and $\sum_j \epsilon_j^2 = -2\sum_{j<k}\epsilon_j\epsilon_k$, which follows from the constraint $\sum_j \epsilon_j = 0$. The integral $\int \frac{dq}{2\pi}\frac{1}{|q|} \cos[q(x_j-x_k)] = \text{const.} -(1/\pi)\ln x$ then gives the exponent $-\frac{2}{K} \sum_{j < k} \epsilon_j \epsilon_k \ln|x_j-x_k|$, which reproduces Eq.~\eqref{eq:Zdw}.  

\section{Correlations in \texorpdfstring{$\ket{\psi_{\text{n.c.}}}$}{psi nc}}\label{sec:nccorrelations}

In this Appendix we discuss correlations in $\ket{\psi_{\text{n.c.}}}$, which are relevant to the transition in Sec.~\ref{sec:qnd}. Using the domain-wall description in Appendix~\ref{sec:walls} we can discuss correlation functions in the state $\ket{\psi_{\text{n.c.}}}$ for $K<1$ as well as $K>1$. First we consider correlations of the smooth part of the particle density. These are computed in $\ket{\psi_{\text{n.c.}}}$ as
\begin{align}
	\braket{\nabla \hat \phi(0) \nabla \hat \phi(x)}_{\text{n.c.}} &= \frac{\text{Tr}[e^{-\beta \hat H} \hat M^2_{\text{n.c.}}  \nabla \hat \phi(0) \nabla \hat \phi(x)] }{\text{Tr}[e^{-\beta \hat H} \hat M^2_{\text{n.c.}} ]} = \frac{\int D\varphi \nabla \varphi(0) \nabla \varphi(x) e^{-s_{\text{n.c.}}[\varphi]} }{ \int D\varphi e^{-s_{\text{n.c.}}[\varphi]}},
\end{align}
where in the first line we have used $[\hat M_{\text{n.c.}},\nabla \hat \phi]=0$. The field $\varphi(x)$ describes density fluctuations at a fixed imaginary time. The action $s_{\text{n.c.}}$ is given in Eq.~\eqref{eq:snc}. For $K>1$ the $\cos 2\varphi$ term is irrelevant under RG, and so long-wavelength correlations can be computed with respect to $s[\varphi]$ in Eq.~\eqref{eq:smalls}. Dimensional analysis immediately reveals that in this regime $\braket{\nabla \hat \phi(0) \nabla \hat \phi(x)}_{\text{n.c.}} \sim x^{-2}$. For $K<1$ the $\cos 2\varphi$ term is relevant, and then at long wavelengths the domain-wall description is appropriate.

For $K<1$ we use the approximation of sharp domain walls $\nabla \varphi(x) \simeq \pi \sum_j \epsilon_j \delta(x-x_j)$. With this parametrization, $\braket{\nabla \hat \phi(0) \nabla \hat \phi(x)}_{\text{n.c.}}$ is in this theory a correlation function for the locations of domain walls. Expanding Eq.~\eqref{eq:Zdw} in powers of $g$ we find that at $O(g^2)$,
\begin{align}
	\braket{\nabla \hat \phi(0) \nabla \hat \phi(x)}_{\text{n.c.}} \simeq - g^2 x^{-2/K}.
\end{align}
This is the contribution to the correlation function from the saddle point featuring two oppositely signed domain walls at locations $0$ and $x$. We can also ask about the contribution from quadratic fluctuations around a given saddle point. To do this for the saddle point with no domain walls, we expand $\cos[2\varphi]$ to generate a mass for the field $\varphi$. Alone, this term describes short-range correlations, and it is straightforward to show that if we treat $s[\varphi]$ perturbatively the contribution to $\braket{\nabla \hat \phi(0) \nabla \hat \phi(x)}_{\text{n.c.}}$ scales as $x^{-4}$. Therefore for $1/2 < K < 1$ the asymptotic behavior is $x^{-2/K}$. These results show that correlations between density fluctuations decay more rapidly in space that in the unmeasured state. 

The description in terms of domain walls also allows us to calculate phase correlations, and here we focus on
\begin{align}
	\braket{e^{i[\hat \theta(x)-\hat \theta(0)]}}_{\text{n.c.}} &= \frac{\int D\theta \text{Tr}[  e^{-\beta \hat H} \hat M_{\text{n.c.}} \ket{\theta} e^{i[\theta(x)-\theta(0)]} \bra{\theta}\hat M_{\text{n.c.}}]}{\int D\theta \text{Tr}[  e^{-\beta \hat H} \hat M_{\text{n.c.}} \ket{\theta} \bra{\theta}\hat M_{\text{n.c.}}]}.\notag \\
	&=\frac{ \int D \theta D\varphi D\varphi' e^{-\frac{1}{2}(s_{\text{n.c.}}[\varphi]+s_{\text{n.c.}}[\varphi']) + \frac{i}{\pi} \int dx' \nabla \theta [\varphi-\varphi']+ i\int_0^x dx' \nabla \theta }}{\int D \theta D\varphi D\varphi' e^{-\frac{1}{2}(s_{\text{n.c.}}[\varphi]+s_{\text{n.c.}}[\varphi']) + \frac{i}{\pi} \int dx' \nabla \theta [\varphi-\varphi']}}.\label{eq:thetatheta}
\end{align}
Here the first equality follows from inserting a resolution of the identity in the basis of $\hat \theta$ eigenstates, and the second follows from $\braket{\varphi|\theta} = e^{\frac{i}{\pi} \int dx \nabla \theta \varphi}$, where $\ket{\varphi}$ is a $\hat \phi$ eigenstate. Note that here it has been necessary to introduce two fields $\varphi$ and $\varphi'$ because $[\hat M_{\text{n.c.}},e^{i \hat \theta}] \neq 0$, and that a factor $1/2$ appears before each of $s_{\text{n.c.}}[\varphi]$ and $s_{\text{n.c.}}[\varphi']$. This factor is a consequence of the fact that, for example, $s_{\text{n.c.}}[\varphi]$ is determined by the integration over fluctuations of $\phi(x,\tau)$ for $\tau > 0$ only. If we integrate out the $\theta$ field we enforce
\begin{align}
	\varphi'(x') = \varphi(x') + \pi T_{0,x}(x')
\end{align}
in the numerator, where $T_{0,x}(x')=1$ for $0 \leq x' \leq x$ and $T_{0,x}(x')=0$ otherwise. In the denominator we instead have $\varphi'=\varphi$. We see then that the expectation value $\braket{e^{i[\hat \theta(x)-\hat \theta(0)]}}_{\text{n.c.}}$ is the ratio of two partition functions: in the denominator $\varphi$ and $\varphi'$ are forced to be equal to one another across all of space, while in the numerator they differ by $\pi$ in the interval $[0,x]$.

Although it is certainly not the simplest approach, it will be instructive to see how the behavior $\braket{e^{i[\hat \theta(x)-\hat \theta(0)]}}_{\text{n.c.}} \sim x^{-1/(2K)}$ for $K>1$ arises from Eq.~\eqref{eq:thetatheta}. In this regime measurements are irrelevant, so we consider the case where $v=0$. Then
\begin{align}
	\braket{e^{i[\hat \theta(x)-\hat \theta(0)]}}_{\text{n.c.}} &= \frac{\int D\varphi \, e^{-\frac{1}{2}(s[\varphi]+s[\varphi+T_{0,x}])}}{\int D\varphi \, e^{-s[\varphi]}}, \notag \\
	\frac{1}{2}(s[\varphi]+s[\varphi+T_{0,x}]) = s[\varphi] &+ \frac{1}{K} \int \frac{dq}{2\pi} |q| \tilde \varphi(q) \tilde T_{0,x}(-q) +  \frac{\pi}{2 K} \int \frac{dq}{2\pi} |q| |\tilde T_{0,x}(q)|^2.
\label{eq:actionwithT0y}
\end{align}
Integrating out $\varphi$ then leads to
\begin{align}
	\braket{e^{i[\hat \theta(x)-\hat \theta(0)]}}_{\text{n.c.}} = e^{-\frac{\pi}{4K} \int \frac{dq}{2\pi} |q| |\tilde T_{0,x}(q)|^2}. \label{eq:thetacorrelatorlargeK}
\end{align}
The integral $\int_{-\Lambda}^{\Lambda} \frac{dq}{2\pi} |q| |\tilde T_{0,x}(q)|^2 = 4 \int_{-\Lambda}^{\Lambda} \frac{dq}{2\pi} |q|^{-1} \sin^2 (qx/2) \simeq \frac{2}{\pi} \ln x$ up to an additive constant, and from this we find the decay $\braket{e^{i[\hat \theta(x)-\hat \theta(0)]}}_{\text{n.c.}} \sim x^{-1/(2K)}$. 

For $K<1$ the measurements $v$ dominate on the largest scales. For the leading saddle point, corresponding to no domain walls, we expand $\cos2\varphi \simeq 1-2\varphi^2$ and then integrate out $\varphi$. For $x \gg v^{-1}$ the behavior is qualitatively similar to setting $\tilde \varphi(q) \sim \delta(q)$, which gives
\begin{align}
	\frac{1}{2}(s_{\text{n.c.}}[\varphi]+s_{\text{n.c.}}[\varphi+T_{0,x}]) &= \frac{\pi}{2 K} \int \frac{dq}{2\pi} |q| |\tilde T_{0,x}(q)|^2, \notag\\ 
	\braket{e^{i[\hat \theta(x)-\hat \theta(0)]}}_{\text{n.c.}} &= e^{-\frac{\pi}{2K} \int \frac{dq}{2\pi} |q| |\tilde T_{0,x}(q)|^2} \sim x^{-1/K}, \label{eq:B7}
\end{align}
to be contrasted with $x^{-1/(2K)}$ for $K>1$. Although phase correlations remain algebraic, their decay is significantly faster when measurements are relevant.

\section{Correlations in the ensemble of $\ket{\psi_m}$}\label{sec:mcorrelations}
Here we discuss the calculations of the correlation functions $C_N(x)$ and $D_N(x)$ in Sec.~\ref{sec:main}. For $K>1/2$ the measurements are irrelevant, and as a consequence the long-wavelength behavior of these correlation functions can be understood by considering perturbations around the unmeasured system. For $K<1/2$ the replicas are locked together by the measurements, and we can compute $C_N(x)$ and $D_N(x)$ in a similar way to $\braket{\nabla \hat \phi(0) \nabla \hat \phi(x)}_{\text{n.c.}}$ and $\braket{e^{i[\hat \theta(x)-\hat \theta(0)]}}_{\text{n.c.}}$, respectively.
 
First note that we can write $C_2(x)$ as
\begin{align}
	C_2(x) = \frac{\int \prod_{\alpha} D\varphi_{\alpha}   \frac{1}{2} \nabla \bar{\varphi}_{\pi}(0) \nabla \bar{\varphi}_{\pi}(x) e^{-s_{2}[\{\bar{\varphi}_{\kappa}\}]} }{\int \prod_{\alpha} D\varphi_{\alpha} e^{-s_{2}[\{\bar{\varphi}_{\kappa}\}]}},
\end{align}
For $N=2$ we can use the long-wavelength action in Eq.~\eqref{eq:spm} and we immediately recognize a variant of the action in Eq.~\eqref{eq:snc} describing fluctuations of the antisymmetric field $\bar{\varphi}_{\pi}$, albeit with a modified Luttinger parameter of $2K$. From the domain-wall description used in Sec.~\ref{sec:nccorrelations} we find
\begin{align}
	C_2(x) \sim -x^{-1/K}
\end{align}
at long wavelengths. The phase correlations $D_2(x)$ can be computed similarly. First note that for integer $N \geq 2$ we have
\begin{align}
	D_N(x) &= \frac{\int Dm D\theta_0 D\theta_1 \text{Tr}[  e^{-\beta \hat H} \hat M_m \ket{\theta_0,\theta_1} e^{i[\theta_0(x)-\theta_0(0)]}e^{-i[\theta_1(x)-\theta_1(0)]} \bra{\theta_0,\theta_1} \hat M_m]}{\int Dm \text{Tr}[  e^{-\beta \hat H} \hat M^2_m] }.\notag \\
	&=\frac{ \int Dm \prod_{\alpha=0}^{N-1} D \theta_{\alpha}  D\varphi_{\alpha} D\varphi'_{\alpha} e^{-\frac{1}{2} \sum_{\alpha}(s_{\mu}[\varphi_{\alpha},m]+s_{\mu}[\varphi'_{\alpha},m]) + \frac{i}{\pi} \int dx' \nabla \theta_{\alpha} [\varphi_{\alpha}-\varphi'_{\alpha}]+ i\int_0^x dx' (\nabla \theta_0-\nabla \theta_1) }}{ \int Dm \prod_{\alpha=0}^{N-1} D\varphi_{\alpha}  e^{-s_{\mu}[\varphi_{\alpha},m]}}.\label{eq:thetathetareplica}
\end{align}
In the interest of brevity, in the numerator of the second line we have introduced resolutions of the identity in the $\theta$ basis for every replica, although these were only necessary for $\alpha=0,1$. Integrating out $\theta_{\alpha}$ for $\alpha \geq 2$ we fix $\varphi'_{\alpha}=\varphi_{\alpha}$, whereas $\varphi'_{0} = \varphi_{0} + \pi T_{0,x}$ and  $\varphi'_{1} = \varphi_{1} - \pi T_{0,x}$. On integrating out the measurements $m$ we couple the different $\varphi_{\alpha}$ and the result has the form
\begin{align}
	D_N(x) = \frac{ \prod_{\alpha=0}^{N-1} D\varphi_{\alpha} e^{-s_N[\{\varphi_{\alpha}\}] - \frac{1}{K} \int \frac{dq}{2\pi} |q| (\tilde \varphi_{0}(q) - \tilde \varphi_{1}(q))\tilde T_{0,x}(-q) - \frac{\pi}{K}\int \frac{dq}{2\pi}|q| |T_{0,x}(q)|^2 + \ldots }}{ \prod_{\alpha=0}^{N-1} D\varphi_{\alpha} e^{-s_N[\{\varphi_{\alpha}\}]}},\label{eq:thetathetareplica2}
\end{align}
where the ellipsis in the exponent of the numerator represents contributions that are local to $0$ and to $x$. These do not affect the asymptotic behavior of the correlation function. When measurements are irrelevant as for $K>1/2$, at long wavelengths $s_N[\{\varphi_{\alpha}\}]$ behaves as the sum of $N$ Gaussian actions $s[\varphi_{\alpha}]$. We can then integrate out the fields $\varphi_{\alpha}$, and recover the result expected without measurements $D_N(x) \sim x^{-1/K}$. For $K<1/2$ where measurements are relevant, we expect that at long wavelengths the fields $\varphi_{\alpha}$ are locked to one another. In the case $N=2$, transforming variables to $\bar{\varphi}_0$ and $\bar{\varphi}_{\pi}$, we find that fluctuations of the field $\bar{\varphi}_{\pi}$ behave as for $\varphi$ in the no click scenario (there for $K<1$), although comparing the prefactor of $ |T_{0,x}(q)|^2$ in Eq.~\eqref{eq:thetathetareplica2} with that in Eq.~\eqref{eq:actionwithT0y} we see that the effective Luttinger parameter is here $K/2$. This leads to $D_2(x) \sim x^{-2/K}$ for $K<1/2$, to be compared with $\braket{e^{i[\hat \theta(x)-\hat \theta(0)]}}_{\text{n.c.}} \sim x^{-1/K}$ in Eq.~\eqref{eq:B7}.

\section{Numerical results}\label{sec:numerics}
Here we present numerical results on correlation functions in states $\ket{\psi_{\text{n.c.}}}$ discussed in Sec.~\ref{sec:qnd}. Our focus is on the $XXZ$ spin chain $\hat H_{XXZ} = \sum_{j} [\hat S^x_j \hat S^x_{j+1} + \hat S^y_j \hat S^y_{j+1} + \Delta \hat S^z_j \hat S^z_{j+1}]$ in the sector with $\sum_j \hat S^z_j=0$. Through a Jordan-Wigner transformation this model is equivalent to spinless fermions with nearest-neighbour density interactions and at half filling. For $|\Delta|<1$ the ground states of this model are critical, and at long wavelengths the behavior is then described by TLL theory with Luttinger parameter $K$ given by $\Delta = -\cos[\pi/(2K)]$ \cite{giamarchi2003quantum}. For $\Delta=0$ and hence $K=1$, the model describes non-interacting fermions. The state of interest here is
\begin{align}
	\ket{\psi}_{\text{n.c.}} \propto e^{-V \sum_{j=0,2,\ldots} \hat S^z_j} \ket{\psi}_{\text{g.s.}},
\label{eq:Mnc_microscopic}
\end{align}
and the quantity $V$ differs from the parameter $v$ in Eq.~\eqref{eq:snc} only by a constant of order unity. Our approach is to prepare approximate ground states $\ket{\psi}_{\text{g.s.}}$ of $\hat H_{XXZ}$ for various $K$ using iDMRG methods from the TenPy library \cite{tenpy}. Naturally there are limitations in using this method to prepare critical states, for example the use of a finite bond dimension gives rise to a finite correlation length $\xi \propto \chi^{\kappa}$ with $\kappa \simeq 1.3$ \cite{pollmann2009theory}. In practice, convergence is poor for $K$ substantially below unity, where there are strong density correlations, and so we restrict ourselves to $K \geq 4/5$. The iDMRG algorithm prepares a MPS representation of $\ket{\psi}_{\text{g.s.}}$ with a unit cell of two sites $j=0,1$. Clearly $\ket{\psi}_{\text{n.c.}}$ can be represented by a MPS with the same periodicity, so we can prepare its MPS representation from that of $\ket{\psi}_{\text{g.s.}}$ simply by acting with $e^{-V S^z_{j=0}}$ and normalizing the result. For $K=1$ ($\Delta=0$) there is the additional possibility of performing exact numerical calculations using fermionic Gaussian states, and so in this case we can compare the two approaches.

\begin{figure}
	\includegraphics[width=\textwidth]{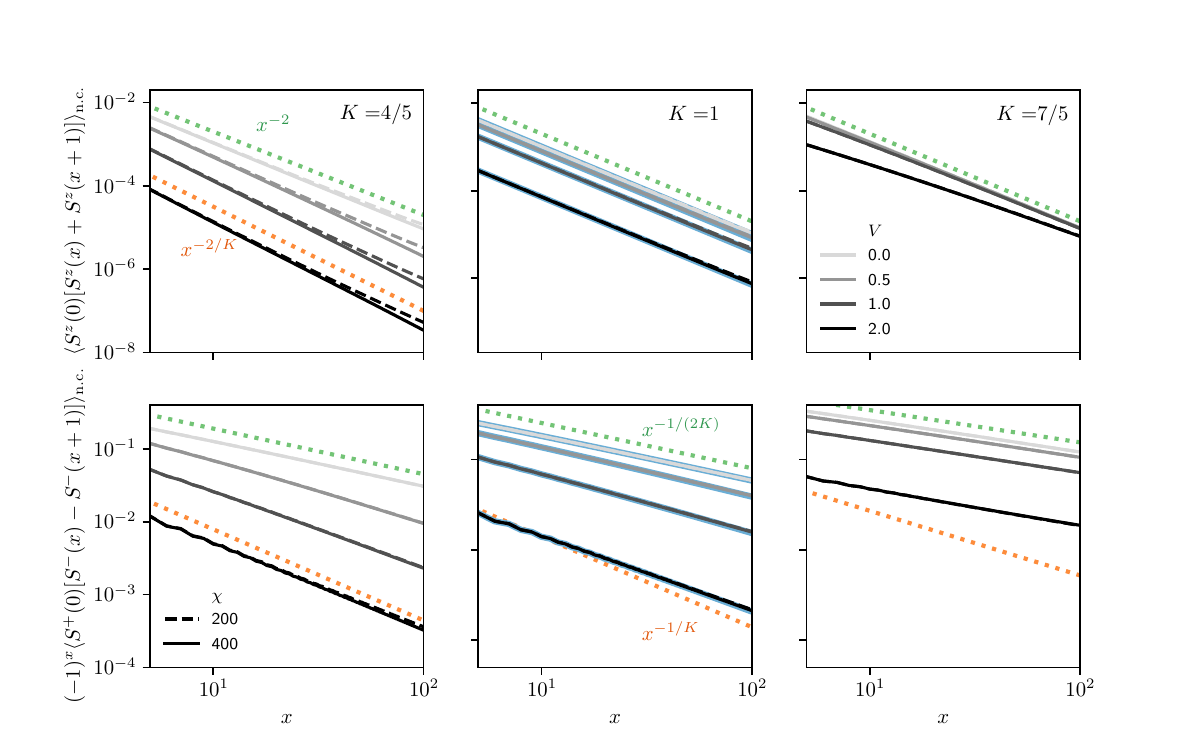}
\caption{Correlation functions computed in $\ket{\psi_{\text{n.c.}}}$ for the $XXZ$ model using iDMRG. The upper and lower panels show respectively the $\nabla \hat \phi$ and $e^{i\hat \theta}$ correlation functions, and the Luttinger parameters $K$ vary from column to column. The different bond dimensions $\chi$ (shown dashed and solid) and measurement strengths (shades) are indicated on the legends, and are the same for all panels. Dotted lines show theoretical predictions: in the upper right panel the orange line shows the behavior $x^{-2/K}$, while green lines in the upper panels show $x^{-2}$. In the lower panels orange lines show $x^{-1/K}$, while green lines show $x^{-1/(2K)}$ (see text for details). In the central panels we compare iDMRG results (greys) with exact results for a system of $L=3 \times 10^3$ sites with periodic boundary conditions (blue, $V$ increasing from top to bottom).}
\label{fig:DMRG}
\end{figure}

Two correlation functions that change their functional dependence on $K$ across the transition at $K=1$ are $\braket{\nabla \hat \phi(0) \nabla \hat \phi(x)}_{\text{n.c.}}$ and $\braket{e^{i[\hat \theta(x)-\hat \theta(0)]}}_{\text{n.c.}}$. The first of these describes the smooth part of the particle density, while the second is a phase correlator. As we have shown in Appendix~\ref{sec:nccorrelations}, the former decays as $-x^{-2}$ for $K>1$ and as $-x^{-2/K}$ for $K<1$, while the latter decays as $x^{-1/(2K)}$ for $K>1$ and as $x^{-1/K}$ for $K<1$. To relate these to correlation functions of the spins we write 
\begin{align}
	S^z_j \simeq -\pi^{-1} \nabla \hat \phi(x) + \pi^{-1} (-1)^x \cos[2\hat \phi(x)], \quad S^+_j \simeq (2\pi)^{-1/2} e^{-i\theta(x)}\big[(-1)^x + \cos [2\hat\phi(x)]\big],
\end{align}
where $x$ is the continuum analogue of the site $j$. Then, at large $x$,
\begin{align}
	\braket{S^z_0 [ S^z_j + S^z_{j+1}]} \sim \braket{\nabla \hat \phi(0) \nabla \hat \phi(x)}_{\text{n.c.}},  \quad (-1)^j \braket{S^+_0 [ S^-_j - S^-_{j+1}]} \sim \braket{e^{i(\theta(x)-\theta(0))}}_{\text{n.c.}}, \label{eq:spincorrelators}
\end{align}
where we have omitted prefactors, and have additionally neglected contributions to the right-hand sides of these relations that decay more rapidly with $x$ than those displayed. For brevity we here refer to the correlators on the left hand side, that are defined at the lattice scale and straightforward to calculate numerically, as the $\nabla \hat \phi$ and $e^{i\hat \theta}$ correlators, respectively. Because our weak measurements act on the even sites, we restrict the $\nabla \hat \phi$ correlator to odd values of $j$; results for even $j$ are qualitatively similar but there is a $V$-dependent offset relative to odd $j$. We make no such restriction for the $e^{i\hat \theta}$ correlator. We show numerical results for these correlation functions in Fig.~\ref{fig:DMRG}. As noted above, for $K=1$ the $XXZ$ model corresponds to free fermions, so we also show results from calculations based on fermionic Gaussian states.

First note that with $V=0$, in which case there is no measurement and $\ket{\psi_{\text{n.c.}}}=\ket{\psi_{\text{g.s.}}}$, the power-law decays of the $\nabla \hat \phi$ and $e^{i\hat \theta}$ correlators Eq.~\eqref{eq:spincorrelators} indeed match TLL theory, decaying respectively as $x^{-2}$ (upper panels) and $x^{-1/(2K)}$ (lower panels). Additionally, for $K=1$ where the measurement is a marginal perturbation, the $\nabla \hat \phi$ correlator is simply rescaled: For small $V$ the leading order contribution arises at second order in perturbation theory, taking the form $V^2 x^{-2}$. The behavior of the $e^{i\hat \theta}$ correlation function at $K=1$ is more difficult to ascertain; our theory predicts a sharp jump from $x^{-1/(2K)}$ to $x^{-1/K}$ as $K$ is decreased through unity, but this jump is smoothed out for finite $L$ (as in the results from exact diagonalization) and for finite $\chi$ (as in the iDMRG calculations). Our focus here is on behavior in the two phases, and so we defer discussion of the critical point $K=1$ to future work.

We now discuss the behavior of the $\nabla \hat \phi$ correlation function for $K=4/5$ and for $K=7/5$. For $K=4/5$ the expected change in the exponent from $x^{-2}$ to $x^{-2/K}=x^{-5/2}$ is not straightforward to observe on these scales; it is nevertheless clear that the measurement-induced change in this correlation function is far more significant at smaller $K$. Most striking is the fact that, for $K=7/4$ and a measurement so strong as $V=1$, the $\nabla \hat \phi$ correlation function is essentially unchanged, while even for $V=2$ it clearly approaches its unperturbed value as $x$ is increased. This is precisely the behavior expected for measurements that are irrelevant in the RG sense. 

The $e^{i\hat \theta}$ correlation function shows stronger signatures of the transition. For $K=4/5$ we expect for $V=0$ a slow decay $x^{-5/8}$, while for $V \neq 0$ we expect $x^{-5/4}$ provided we go to sufficiently large $x$. Note that the length scales required to observe this crossover diverge as $V \to 0$, so it is unsurprising that the behavior $x^{-5/8}$ is only observed for the larger values of $V$. For $K=7/5$ on the other hand, although our density measurements suppress the prefactor in the phase correlations, there is as expected no visible change in the power of the decay. In summary, the results of this section demonstrate a sharp contrast in the structure of $\ket{\psi_{\text{n.c.}}}$ for $K<1$ relative to $K>1$.

The numerical calculations in this section are relevant for the transition at $K=1$ studied in Sec.~\ref{sec:qnd}, and it is natural to ask whether similar results can be obtained for the generic transition at $K=1/2$ discussed in Sec.~\ref{sec:main}. A barrier to doing this in the $XXZ$ spin chain above is that decreasing the Luttinger parameter toward $K=1/2$ corresponds to increasing the anisotropy parameter toward the Heisenberg point $\Delta=1$. For $\Delta>1$ there is a quantum phase transition from the gapless phase of interest into a phase with long-range antiferromagnetic order. In language appropriate for spinless fermions, this corresponds to CDW order induced by strong repulsive interactions. To study $K<1/2$ there are a number of possibilities, for example one could include next-nearest-neighbor interactions, or even long-range interactions, and thereby frustrate the order that would otherwise set in for $\Delta>1$. Given a lattice model which exhibits TLL behavior for $K<1/2$, it is then necessary to ensure that the slow decay of density correlations $x^{-2K} < x^{-1}$ is not cut off by bond dimension truncation. Such a calculation will be essential to observe the $K=1/2$ transition using the approach outlined in Sec.~\ref{sec:postselection}.

\section{Gaussian measurements}\label{sec:gaussian}

As discussed in Sec.~\ref{sec:main}, in constructing our replica field theory it is convenient to use a measurement model of the form Eq.~\eqref{eq:Mm}. In this Appendix we discuss how $\hat M_m$ can be implemented by coupling local densities to QHOs. For simplicity we will start from a lattice model of spinless fermions at half filling so that the normal-ordered density operator $\hat n(x)$ at site $x$ has the property $\hat n^2(x)=1/4$. Note that this choice causes the $n^2(x)$ contribution to the action $s_{\mu}[\varphi,m]$, relevant to the discussion below Eq.~\eqref{eq:individual1}, to cancel between numerator and denominator in the calculation of postmeasurement expectation values.

Let us first write the many-body state in the basis of density eigenstates $\ket{\psi_{\text{g.s.}}} = \sum_n \braket{n|\psi_{\text{g.s.}}}\ket{n}$, where $\hat n(x)\ket{n}=n(x)\ket{n}$ for $n(x)=\pm 1/2$. At each $x$ we introduce an oscillator, and we denote their position operators by $\hat m(x)$. The oscillators are taken to have frequencies $\omega$, and `masses' of $\mu/\omega$ so that $\mu$ is the inverse-square oscillator length. The Hamiltonian of the QHO at $x$ should take the form
\begin{align}
	\hat H_x(t) = \frac{\omega}{2\mu} \hat \pi^2(x) + \frac{1}{2}\mu \omega \hat m^2(x) - f(t)\hat m(x) \hat n(x),
\end{align}
where $\hat \pi(x)$ is the momentum, with $[\hat m(x),\hat \pi(x)]=i$. At $t=0$ we suppose that $f(0)=0$ and that the QHOs are in the corresponding ground state $\ket{\Omega}$. The initial state of the system and QHOs is simply the tensor product $\ket{\Psi(t=0)} = \ket{\psi_{\text{g.s.}}} \otimes \ket{\Omega}$, and so has amplitudes
\begin{align}
	\braket{n|\psi_{\text{g.s.}}}\braket{m|\Omega} \propto \braket{n|\psi_{\text{g.s.}}} e^{-\sum_x \frac{1}{4}\mu m^2(x)}
\end{align}
From $t=0$ to $t=T$ we increase the coupling to $f(T)=\mu \omega$ at every $x$ so that
\begin{align}
	\hat H_x(T) =  \frac{\omega}{2\mu} \hat \pi^2(x) + \frac{1}{2}\mu \omega[ \hat m(x) - \hat n(x)]^2 - \frac{1}{2}\mu \omega \hat n^2(x).
\end{align}
If this increase is adiabatic with respect to the oscillator and sudden with respect to the system the amplitudes become
\begin{align}
	\braket{n,m|\Psi(T)} \propto \ket{\psi_{\text{g.s.}}} e^{- \frac{1}{4}\mu\sum_x [m(x)-n(x)]^2}.
\end{align}
Performing now a projective measurement of the QHO with result $m$ generates a state $\ket{\psi_m}$ with amplitudes
\begin{align}
	\braket{n|\psi_m} \propto  e^{- \frac{1}{4}\mu \sum_x[m(x)-n(x)]^2}\braket{n|\psi_{\text{g.s.}}}.
\end{align}
If we now take the continuum limit we find the operation represented by $\hat M_m$ in Eq.~\eqref{eq:Mm}. For the coupling between the QHO and the system to be adiabatic with respect to the QHO, we require $T \gg \omega^{-1}$. On the other hand, for it to be sudden with respect to the system we require $T \ll \Lambda^{-1}$, where $\Lambda$ is the UV cutoff. This implies $\omega \gg \Lambda$. In order that our measurement is weak, we also require $\mu^{-1}$ to be large relative to the variance of the particle density. Large $\omega$ and small $\mu$ imply a small mass for the QHO.

\twocolumngrid

%

\end{document}